\documentclass[10pt,letterpaper]{article}
\usepackage[top=0.85in,left=2.75in,footskip=0.75in]{geometry}

\usepackage{amsmath,amssymb}
\usepackage{algorithm,algpseudocode}

\algnewcommand\algorithmicforeach{\textbf{for each}}
\algdef{S}[FOR]{ForEach}[1]{\algorithmicforeach\ #1\ \algorithmicdo}

\usepackage{changepage}

\usepackage{textcomp,marvosym}

\usepackage{cite}

\usepackage{nameref,hyperref}

\usepackage[right]{lineno}

\usepackage[nopatch=eqnum]{microtype}
\DisableLigatures[f]{encoding = *, family = * }

\usepackage[table]{xcolor}

\usepackage{array}

\newcolumntype{+}{!{\vrule width 2pt}}

\newlength\savedwidth

\newcommand\thickhline{\noalign{\global\savedwidth\arrayrulewidth\global\arrayrulewidth 2pt}%
\hline
\noalign{\global\arrayrulewidth\savedwidth}}

\raggedright
\usepackage[aboveskip=1pt,labelfont=bf,labelsep=period,justification=raggedright,singlelinecheck=off]{caption}

\makeatletter
\renewcommand{\@biblabel}[1]{\quad#1.}
\makeatother

\usepackage{lastpage,fancyhdr,graphicx, subcaption}
\usepackage{epstopdf}
\fancyheadoffset[L]{2.25in}
\fancyfootoffset[L]{2.25in}
\newcommand{\maxinfo}{{\max_{\text{info}}}}

\usepackage{amssymb}
\usepackage{pifont}
\newcommand{\cmark}{\ding{51}}
\newcommand{\xmark}{\ding{55}}

\usepackage[table]{xcolor}
\usepackage{multirow}
\usepackage{booktabs}

\begin{document}
\vspace*{0.2in}

\begin{flushleft}
{\Large
	\textbf\newline{Algorithmic Information Theory for Graph Edge Grouping and Substructure Analysis} 
}
\newline
\\
Gabriel Potestades\textsuperscript{1\textcurrency *}
\\
\bigskip
\textbf{1} College of Computer Studies/Human-X Interactions Lab, De La Salle University, Manila City, Metro Manila, Philippines
\\
\bigskip

\textcurrency  Current Address: College of Computer Studies/Human-X Interactions Lab, De La Salle University, Manila City, Metro Manila, Philippines \\
\bigskip
* \text{gabriel\_potestades@dlsu.edu.ph}

\end{flushleft}
\section*{Abstract}
Understanding natural phenomenon through the interactions of different complex systems has become an increasing
focus in scientific inquiry. Defining complexity and actually measuring it is an ongoing debate and no standard 
framework has been established that is both theoretically sound and computationally practical to use.
Currently, one of the fields which attempts to formally define complexity is in the realm of Algorithmic
Information Theory. The field has shown advances by studying the complexity values of binary strings and 
2-dimensional binary matrices using 1-dimensional and 2-dimensional Turing machines, respectively. 
Using these complexity values, an algorithm called the Block Decomposition Method developed by Zenil, et al. in
2018, has been created to approximate the complexity of adjacency matrices of graphs which have found relative
success in grouping graphs based on their complexity values. We use this method along with another method 
called edge perturbation to exhaustively determine if an edge can be identified to connect two subgraphs 
within a graph using the entire symmetric group of its vertices permutation and via unique permutations we call
automorphic subsets, which are a special subset of the symmetric group. We also analyze if edges will be 
grouped closer to their respective subgraphs in terms of the average algorithmic information contribution. 
This analysis ascertains if Algorithmic Information Theory can serve as a viable theory for understanding graph 
substructures and as a foundation for frameworks measuring and analyzing complexity. The study found that the 
connecting edge was successfully identified as having the highest average information contribution in 29 out of
30 graphs, and in 16 of these, the distance to the next edge was greater than $\log_2(2)$. Furthermore, the 
symmetric group ($S_G$) outperformed automorphic subsets in edge grouping.

\section*{Author summary}
This work reports on the empirical result of the application of the aforementioned complexity values 
particularly for 2-dimensional binary matrices. By retrieving the complexity of a graph's matrix using the 
Block Decomposition Method, we can determine the algorithmic information contribution of an edge before and 
after it was removed from the graph via a method called edge perturbation. We perform averaging of algorithmic 
information contribution of edges in graph with respect to different permutations. Since the symmetric group of
a graph with $n$ nodes grows by $n!$, runtime also becomes long and thus graphs between 9 and 12 nodes only 
were analyzed.


\section*{Introduction}
Complexity has been a topic of interest in science since the 1970s and has sprouted a field called
Complex Systems. The field studies the properties and patterns found in a collection of entities interacting 
with one another within a system~\cite{bib1}. One way of analyzing or understanding complex systems is 
through networks. Networks are ubiquitous in nature, such as animal social networks, citation networks and 
social networks~\cite{bib2}. They provide us with a relational view of the structure of interaction of
entities in a complex system. Although we use networks to analyze complex systems, the very notion of a
complex system is not concrete and no universal definition has been agreed upon in the scientific community.

Finding a clear-cut definition of what constitutes a complex system has been disputed for years 
now~\cite{bib3}. By nature of complex systems, finding a compressed and simple description of how these systems
behave is a difficult task. Since one of the primary objectives of science is to understand nature and reality 
using unified and simple explanations, defining complexity is embedded in this objective and involves a zoomed 
out lens view on explaining nature rather than a reductionist approach. By creating a definition of what a 
complex system is, we can properly analyze these systems that can be found in nature.

In our current scientific climate, to explain phenomenon in nature, we employ the use of computational methods.
When we have observational data of a phenomenon, we attempt to find equations / algorithms that explain the 
phenomenon accurately. We consider our computable functions as explanations of why nature behaves in certain 
ways. Thus, computable functions-- any function that can be executed by a Universal Turing Machine~\cite{bib4}, 
are our \textit{compressed} descriptions of understanding nature.

One of the candidates of defining complexity while simultaneously aims to provide a computational explanation
of observed data which involves the shortest possible description is in the field of Algorithmic Information
Theory (AIT). Through the approximations of Kolmogorov-Chaitin complexity, advances in this field have led to 
the development of the Block Decomposition Method (BDM)~\cite{bib5}. This method computes the complexity
of binary strings~\cite{bib6} and networks (via binary adjacency matrices)~\cite{bib7} in a more 
algorithmic and agnostic perspective rooted from choosing the shortest possible computable function that
explains the input data. 

Kolmogorov-Chaitin complexity is ultimately uncomputable due to the fact that this function finds the shortest
possible computer program that outputs the data and thus can be reduced to the undecidability of the Halting 
Problem. Consequently, the Block Decomposition Method is an approximation of the algorithmic complexity of 
binary strings and adjacency matrices. Although they are approximations, it was shown that this method can be
used to remove edges from a graph to identify substructures ~\cite{bib8} via an algorithm named causal 
deconvolution, and it was also shown that this method was used on different networks such as dynamical systems 
reprogram them move away or towards randomness~\cite{bib9} and analysis of ecological networks~\cite{bib18}.

We extend and validate the causal deconvolution algorithm implementation done in~\cite{bib8} by using 
permutations of vertices to relabel them and then perform edge perturbation. Although the causal deconvolution 
algorithm managed to identify edges that connect multiple subgraphs such as $K$-ary trees, small-world 
connected to a complete graphand a graph consisting of a random graph, a complete graph and a star graph, this
was only done for one permutation of the graphs' adjacency matrices which led to the motivation for this work 
of employing combinatorics to have a more thorough application of BDM to the graphs.

We use the canonical labelling algorithm created by McKay and Piperno~\cite{bib10} to group different 
permutations from the symmetric group of a graph's vertices into their respective automorphic subsets. Selected 
permutations per automorphic subset is used to get the average information contribution of each edge to 
determine if they will group with other edges that are within the same sub-graph. Comparison is also made when 
using the entire symmetric group in getting the average information contribution to show which set of
permutations is more effective in grouping edges. 

\section*{Materials and methods}

\subsection*{Algorithmic Information Theory}

Algorithmic Information Theory uses the program size of Turing machines to determine the randomness of 
information. It states that, for a given string, if the size of the shortest program that outputs
the string is greater than or equal to the string itself, then the string is considered random. This theory
formalizes the notion of the complexity of information. 

\subsubsection*{Algorithmic Complexity}
The chief equation to this theory is the Kolmogorov-Chaitin $K_T(s)$ complexity~\cite{bib11,bib12} which is 
defined as:
\begin{eqnarray}
\label{eq:KCcomplexity}
	K_T(s) = \min\{ |p|, T(p)=s \}
\end{eqnarray}

\noindent where it takes a string $s$ and finds the program $p$ that is executable in a (prefix-free) universal
Turing machine $T$, and outputs an integer which is the size of the smallest program that outputs $s$. Although
this complexity measure has been proven to be uncomputable, approximations have been computed to get the 
complexity of strings and binary matrices for practical use~\cite{bib6,bib7}.

\subsubsection*{Algorithmic Probability}
Closely related to algorithmic complexity is the notion of Solomonoff algorithmic probability. This probability
determines how likely a random program will output a piece of information~\cite{bib13}. Algorithmic 
probability is defined as:
\begin{eqnarray}
\label{eq:Sprobability}
	\mathfrak{m}(s) = \sum_{p: T(p) = s} 2^{-|p|}
\end{eqnarray}

\noindent where the probability $\mathfrak{m}(s)$ is the sum of probabilities of all (prefix-free) programs $p$
that are executed by a universal Turing machine $T$ and halt. It can be noted that the shorter the length of 
a program $|p|$, the larger its contribution to the overall summation of program lengths.

\subsubsection*{Coding Theorem Method}

$K_T(s)$ can be approximated using $\mathfrak{m}(s)$ by the smallest program in the summation that generates
$s$ using the universal Turing machine $T$. The Levin Coding Theorem~\cite{bib14,bib15} shows that algorithmic
probability and algorithmic complexity are related:
\begin{eqnarray}
\label{eq:Lcoding}
	K(s) \leq \log \frac{1}{\mathfrak{m}(s)} + c 
\end{eqnarray}

\noindent where $c$ is a constant that is independent of string $s$. This theorem shows that the complexity of 
a string is inversely proportional to its algorithmic probability, meaning that the more probable a string is 
to be generated by a random program, the lower its complexity.

This has theorem been applied to create the Coding Theorem Method (CTM)~\cite{bib6} by enumerating Turing 
machines from shorter to longer ones based on the Busy Beaver function~\cite{bib16}. Thus, we can approximate 
$K(s)$ from $\mathfrak{m}(s)$ using the output distribution of halting Turing machines:
\begin{eqnarray}
\label{eq:BB}
D(t,k)(s) = \frac{|\{T \in (n,m) : T(p) = s\}|}{|\{T \in (n,m) : T \text{ halts}\}|}
\end{eqnarray}

We can assign a probability value to a string $s$ by running the Busy Beaver function for $t$ states with $k$ 
symbols (0,1 for binary sequences). Instead of counting the number of $1$s produced by a Turing machine, we
count the number of Turing machines that produce $s$ over the number of all Turing machines for $t$ states and
$k$ symbols that halt.

With this method, we can use $D(t,k)(s)$ as an approximation of $\mathfrak{m}(s)$ and compute algorithmic
complexity:
\begin{eqnarray}
\label{eq:CTM}
CTM(s,t,k) = - \log D(t,k)(s)
\end{eqnarray}

\subsubsection*{Block Decomposition Method}
Since CTM is based on the Busy Beaver function, and the latter being the fastest growing function compared to 
all computable functions, CTM is computationally expensive and ultimately non-computable if to be performed. To
circumvent this problem, BDM was introduced~\cite{bib5} to extend the capabilities of CTM:
\begin{eqnarray}
\label{eq:BDM}
BDM(s,l,m) = \sum_i CTM(s^i, t, k) + \log(n_i) 
\end{eqnarray}

We now fix the values of $t$ and $k$ to have a specific distribution table based on $D(t,k)$, which is 
represented as $CTM$ based in Equation \ref{eq:CTM}. Let $l$ be an integer such that $l \leq s$ and is the 
length of each substring $s^i$ when $s$ is decomposed (with a possible remainder of $y = s \mod l$ characters).
After decomposition, there is a probability that some $s^i$ substrings are the same, thus $n_i$ is the 
multiplicity of each $s^i$ found. If $p_i$ is the smallest program that produces substring $s^i$, 
then $\log(n_i)$ is the number of bits needed to encode $p_i$.

Finally, $m$ is an overlapping parameter to handle possible remainder $y$ when $s$ is not a multiple of $l$. If
$m = l$, then the remaining characters in the string are ignored. If $m < l$, then a sliding window of size $l$
is used and moves $m$ characters until all parts of the string are captured by the window.

\subsubsection*{Block Decomposition Method for 2D matrices}

CTM has also been used in two-dimensional matrices via two-dimensional Turing machines~\cite{bib7}. Instead of 
a one-dimensional tape, the Turing machine's memory is on a two-dimensional grid where up and down is also 
a possible movement of the Turing machine. As a consequence, BDM can also be extended for graphs using their
adjacency matrices and is defined as:
\begin{eqnarray}
\label{eq:BDM2}
BDM_{2D}(X, l) = \sum_{(x_i, n_i) \in X_{(l \times l)}} CTM_{2D}(x_i) + \log(n_i) 
\end{eqnarray}

\noindent where $X$ is an $m \times m$ matrix of a graph with $m$ nodes. The sub-matrix size $l$, where 
$l \leq m$, is used to partition $X$ by $l \times l$ sub-matrices distinctly named $x_i$ per sub-matrix being 
encountered $n_i$ times. Currently, 2D CTM values that were computed were up to $4 \times 4$, which will be 
used in the algorithm. If $l$ is not a multiple of $m$, there will be $r$ rows and $c$ columns that are not 
captured by the partitions, handling of these rows and columns will be also discussed in the algorithm 
implementation.

\subsection*{Graph Theory}
A graph $G$ consists of $(V,E)$ where $V_G$ is a set of $n$ vertices labeled $\{1, 2, \ldots, n \}$ and $E_G$ 
is the set of edges $\{ (u,v) \mid u, v \in V_G\}$ that connect vertices. A graph $G$ with $n$ vertices has a 
symmetric group $S_G$ which is the set of all permutations of $V$. Each permutation $\sigma \in S_G$,
maps a vertex label to another vertex label, $\sigma : V \rightarrow V$, where the edges 
$E^\sigma = \{ (\sigma(u), \sigma(v)) \mid u, v \in V \}$. We can represent the relabeled graph as 
$G^\sigma$. A sub-graph $G_1$ of $G$ is a graph which is defined as $V_{G_1} \subseteq V_G$ and 
$E_{G_1} \subseteq E_G$.

An arbitrary graph $H$ is considered isomorphic to $G$ if there exists a function $\pi: E_G \rightarrow E_H$ 
such that $(u,v) \in E_G$ and $(\pi(u),\pi(v)) \in E_H$ for all $u, v \in E_G$, where $E_G$ and $E_H$ are 
edge sets of each graph respectively.  We denote  $G \simeq H$ to say that both graphs have an isomorphism. 
Each permutation $\sigma \in S_G$ applied to $G$ preserves edge adjacency, thus $G \simeq G^\sigma$. Although 
vertices are relabeled, the structure of the graph is preserved.

An automorphism $\gamma \in S_G$ is also an isomorphism of $G$ (with itself) but with a much stricter 
requirement of  $(\gamma(u), \gamma(v)) \in E \iff (u,v) \in E$, meaning an automorphism not only preserves 
graph structure but also edge adjacency per label of a vertex. Given the original labelling of vertices 
$\{1, 2, \ldots, n \}$, there exists a set of permutations $\{\gamma_1, \gamma_2, \ldots \} \subseteq S_G$ 
that are automorphisms of $G$ which is called the automorphism group $\mathcal{A}(G)$. We use 
$G \overset{\mathcal{A}}{\simeq} H$ to say that two graphs $G, H$ are automorphic.

\subsubsection*{Canonical Labelling}
Canonical labelling was introduced in \cite{bib10} and summarized in \cite{bib17}, it uses vertex degree and
automorphisms to find a definitive relabelling of a graph to test isomorphism with another canonically labelled
graph. The canonical labelling function $C$ relabels the vertices of a graph $G$ such that if $G$ and $H$ 
are isomorphic graphs, then $C(G) = C(H)$.

Let $c_G$ be the permutation of the vertices of the graph $G$ when it is passed to $C$. Using $G$ as the 
original graph, we can apply $\gamma, \sigma \in S_G$, to produce two permuted graphs namely $G^\gamma$ and 
$G^\sigma$. Upon applying $C$ to both permuted graphs, if $c_{G^\gamma} = c_{G^\sigma}$, it means that 
$G^\gamma \overset{\mathcal{A}}{\simeq} G^\sigma$. Since $G^\gamma, G^\sigma$ are just permuted graphs from $G$,
having the same $c$ means that although both graph's nodes are relabeled differently, when $c$ is applied to 
them both, their vertex labels and edges are the same, making them automorphic. We use $c$ to group different 
permutations of $G$ so that each $\sigma \in S_G$ is included in what we call an automorphic subset.

\subsubsection*{Graph Automorphic Subsets}

We can now define an automorphic subset of $G$ as:

\begin{eqnarray}
	\label{eq:AutSub}
	\lambda = 
	\{ \gamma, \sigma, \phi \ldots \ \mid \gamma, \sigma, \phi, \ldots \in S_G
		\land c_{G^\gamma} = c_{G^\sigma} = c_{G^\phi} =  \ldots \}
\end{eqnarray}

Each element in $\lambda$ is a permutation of $V_G$ such that when $V_G$ is permuted by any element in this
set, it is automorphic to any other permutation in the same set. This ultimately groups each permutation 
$\gamma \in S_G$ to a specific automorphic subset. Since a graph can have a number of automorphic subsets, we 
define the complete automorphic set of a graph as:

\begin{eqnarray}
	\label{eq:AutComp}
	\Lambda(G) = \{ \lambda_1, \lambda_2, \lambda_3, \ldots \}
\end{eqnarray}

Each member of $\Lambda(G)$ is a set and also a distinct subset of $S_G$. Note that combining all 
sub-members of all the members of $\Lambda(G)$ gives us $S_G$. The number of automorphic subsets within a 
symmetric group can be determined by the following scenarios:

\begin{itemize}
	\item 1, if the graph is completely symmetric (a complete graph), $\{\mathcal{A}(G)\} = \Lambda(G)$.
	\item $1 < x < |S_G|$, where $x$ is the number of automorphic subsets.
	\item $|S_G|$, if the graph is completely asymmetric (a fully random graph)
\end{itemize}

\subsection*{Algorithms}

If two permutations are automorphic when $G$ is permuted using these permutations, it means that their 
canonical label permutation $c$ is the same. Thus, it can be used to pick an arbitrary permutation $\gamma$ in 
$\lambda$ where $\lambda \in \Lambda(G)$. Pynauty~\cite{bib19} - a Python implementation of the canonical 
labelling package Nauty~\cite{bib10} was used in Algorithm \ref{alg:UnqIso} which retrieves unique 
permutations in $S_G$ so that $G$ can be structurally represented distinctly per automorphic subset. We only 
choose one member from each subset $\lambda \in \Lambda$ since each member of $\lambda$ is already automorphic 
with every other member of the subset thus, one member can represent the entire subset. It is to note that

\begin{eqnarray}
\label{eq:automorphequality}
\Cup(G) = \bigcup_{i=1}^{|\Lambda(G)|} \lambda_i [x]
\end{eqnarray}  

where $i$ is the $i$th member of $\Lambda(G)$ and $[x]$ retrieves an arbitrary member of $\lambda_i$.

\begin{algorithm}
	\caption{Automorphic Subset Sampling}\label{alg:UnqIso}
	\begin{algorithmic}
		\Procedure {$\Cup$}{$G$}
		\State ${\lambda \gets \{\} }$
		\State ${L \gets \{\} }$
		\ForEach {$\gamma \in S_G$}
			\If {$c_{G^\gamma} \not\in L$}
				\State ${\lambda \gets \lambda \cup \{ \gamma \}}$
				\State ${L \gets L \cup \{ c(G^\gamma)\}}$
			\EndIf
		\EndFor
		\State ${\textbf{return } U}$
		\EndProcedure
	\end{algorithmic}
\end{algorithm}

\begin{algorithm}[!t]
	\caption{Average Information Loss via Automorphic Subsets}\label{alg:CausDeco}
	\begin{algorithmic}
		\Procedure {$\mathfrak{C}$}{$G$, $\Gamma$}
			\State ${I \gets \{\}}$ 
			\State ${l \gets 3}$

			\State	
			\State \(\triangleright\) Initialize a set of tuples where the first member of a tuple is an edge 
				of $G$
			\ForEach {$(u,v) \in E_G$}
			\State ${I[(u,v)] \gets 0}$
			\EndFor

			\State
			\ForEach {$\gamma \in \Gamma$}

				\State ${X \gets Adj(E_{G^\gamma})}$
				\State ${\text{bdm}_{X} \gets  BDM_{2D}(X, l)}$
				\State
				\State \(\triangleright\) This loop is where edge perturbation happens

				\ForEach {$(w,x) \in E_{G^\gamma}$}
					\State ${E_{r} \gets E_{G^\gamma} \; \backslash \; (w,x)}$ 
					\State ${Y \gets Adj(E_{r})}$
					\State ${\text{bdm}_Y \gets BDM_{2D}(Y, l)}$
					\State ${\text{bdm}_{(w,x)} \gets \text{bdm}_{X} - \text{bdm}_Y}$
					\State ${(u,v) \gets (\gamma^{-}(w),\gamma^{-}(x))}$
					\State ${I[(u,v)] = I[(u,v)] + \text{bdm}_{(w,x)}}$
				\EndFor
				\State
			\EndFor

			\State
			\State \(\triangleright\) Average all the info loss values using the size of the set of 
				permutations
			\ForEach {$(u,v) \in I$}
				\State ${I[(u,v)] \gets I[(u,v)] \; / \; |\Gamma|}$
			\EndFor
			\State
			
			\State \(\triangleright\) Sort the edges from highest to lowest average info loss
			\State I = sort(I)
			
			\State
			\State ${\textbf{return } I}$
		\EndProcedure
	\end{algorithmic}
\end{algorithm}

Edge perturbation~\cite{bib9} is the process of temporarily removing an edge from a graph to determine how 
much algorithmic information is loss or gained upon removal. To know the information contribution of each edge,
we get the difference of $BDM_{2D}$ values before and after an edge is removed from a graph. Losing information 
implies that the graph is moving towards simplicity while gaining information implies it's moving towards 
randomness. Since an increase in information implies that a longer Turing machine is needed to output the graph
without the edge compared to when it had it.  With this method, we can try and group edges using their 
information contribution values on the basis that $BDM_{2D}$ measures the complexity of a graph. We used 
PyBDM~\cite{bib20}, the Python package that implements $BDM_{2D}$.

$\mathfrak{C}$ (Algorithm \ref{alg:CausDeco}) uses a set of permutations $\Gamma$ to permute the vertices of a 
graph $G$ and perform edge perturbation. Initially, a set of tuples $I$ is created where the average 
algorithmic information contribution of each edge $(u,v) \in E_G$ will be housed and can be retrieved by using 
$(u,v)$ as a key in $I$ which is denoted by $I[(u,v)]$. Using a permutation $\gamma \in \Cup(G)$, we permute 
$G$ and retrieve its adjacency matrix $X$ that is denoted by $Adj(E_{G^\gamma})$. The row and column of the 
matrix are arranged according to the ordered labelling using $V_{G^\gamma}$, where each $v \in V_{G^\gamma}$ is 
the $v$th row/column in the matrix $X$.

For each permuted edge $(w,x) = (\gamma(u),\gamma(v))$ where $u,v \in V_G$, we compute its information
contribution value $\text{bdm}_{(w,x)}$ by removing $(w,x)$ from $G^{\gamma}$ which is denoted by 
$E_{G^\gamma} \; \backslash \; (w,x)$. After removal, the difference in $BDM_{2D}$ between the original matrix 
$X$ and the matrix $Y$ (where the edge is removed) is added to the existing value stored in $I$. To properly 
add the information contribution for the edges when permuted by $\gamma$, we revert $(w,x)$ to its original
label in $E_G$ by the inverse function $\gamma^-$ since calculating the information contribution of each edge
is done in each permutation $\gamma \in \Cup(G)$. Finally, each information contribution value of an edge in 
$I$ is averaged by the number of $\gamma \in \Cup(G)$ since $|\Cup(G)| = |\Lambda(G)|$.

Recalling the definition of $BDM_{2D}$(\ref{eq:BDM2}), $l$ is the size of the 2D $l \times l$ sub-matrices 
produced when $X$ is partitioned. Based on initial testing, it was found that $l = 3$ proved to be more 
effective in grouping edges compared to $l = 4$. We conjecture that since the graphs (9 - 12 nodes) analyzed 
are small, having smaller partitioned sub-matrices is more favorable. We chose the periodic partitioning when
converting a matrix $X$ into blocks of $l \times l$ sub-matrices. This partitioning uses the first 
$|X| \mod l$ rows/columns as padding when the adjacency matrix is not divisible by $l =3$.

Overall, averaging the information contribution of each edge was done to probabilistically account for every
possible unique vertex relabelling. This allows for a more holistic accounting of the algorithmic information 
contribution for each edge to the structure of a graph. 

\subsubsection*{Time Complexity}

The algorithm $\mathfrak{C}$ has a runtime of $\mathcal{O}(n!)$ regardless if $\Gamma = \Cup(G)$ or 
$\Gamma = S_G$ since $\Cup(G)$ iterates over $S_G$ when sampling a permutation for each automorphic subset 
$\lambda \in \Lambda(G)$ for a graph $G$. Inside $\mathfrak{C}$, the edge perturbation algorithm is executed 
and has a linear running time of $\mathcal{O}(m)$, where $m$ is the number of edges in $G$.



\section*{Results}
Edge perturbation was performed on 30 synthetic graphs where each contains two subgraphs \textbf{connected by
a single edge}. We shall label the edge with the highest average information contribution $\maxinfo$ for any of 
the 30 graphs. For probabilistically generated graphs like an Erdős–Rényi graph, a Barabási–Albert graph or a
Watts–Strogatz graph, we used the same graphs for different vertex counts (for each type of graph respectively)
to have a more controlled and rigid experiment of testing edge perturbation. The mentioned graphs have the 
following properties set:

\begin{enumerate}
	\item {Erdős–Rényi - $p = 0.5$, where $p$ is the probability of an edge forming between two vertices.}
	\item {Barabási–Albert - $m = 2$, where $m$ is the number of sampled nodes where a newly added node will
		connect to.}
	\item {Watts–Strogatz - $ p = 0.5$ \& $k = 4$, where $p$ is the probability that an edge will rewire and 
		$k$ is the degree of each vertex.}
\end{enumerate}

We introduce some more notation on the algorithmic complexity of graphs with subgraphs before the discussion 
of results. Let $G$ be a graph that has two subgraphs, $G_1$ and $G_2$ that are connected by a single edge. 
Then the algorithmic complexity of $G$ is

\begin{eqnarray}
\label{eq:algocomplexitytwographs}
K(G) \leq K(G_1) + K(G_2) + O(1)
\end{eqnarray}

where $O(1)$ accounts for the constant program size to connect two vertices from each sub-graph by an edge 
whose program size is independent of $G$. Let $G^*$ be the shortest program that produces $G$ such that 
$K(G) = |G^*|$. If $G$ has two subgraphs, then $|G^*| \leq |G^*_1| + |G^*_2| + O(1)$. When computing 
for $K(G)$, if edges $e_i, e_j \in E_{G_1}$, then the bits needed to encode $e_i$ and $e_j$ into $G$ are 
included in $G_1^*$. There are three main modes of inquiry that can be asked to determine the effectivity of 
edge perturbation using $BDM$:

\begin{itemize}
	\item Does the connecting edge for each graph have a high information contribution value when compared to
		other edges in the same graph?
	\item Do edges that belong to the same sub-graph have similar information contribution values?
	\item Does the connecting edge have a far enough difference from the other edges that it can be identified
		as being produced by a different sub-program in $G$?
\end{itemize}

The following three sections will further expound on these questions respectively.  We compared the average 
information values when using an element in each subset of $\Lambda(G)$ against the symmetric group $S_G$ of a
graph $G$. In practice, computing for the automorphic subsets $\Lambda(G)$ is usually faster (especially for 
regular graphs) than using $S_G$ since $\Cup(G)$ reduces the number of permutations that need to be edge 
perturbed unlike for $S_G$ that considers all permutations of a $G$ which grows $n!$ for $n$ vertices. 

\subsection*{Average Information Contribution of Edges}

We checked if the connecting edge's information contribution is aligned with theoretical expectations in terms
of having a high positive information contribution to a graph. If the information contribution of an edge is 
positive, it implies that information will be lost when the edge is removed. Conversely, if an edge's 
information contribution is negative, it means that the algorithmic complexity of the graph increased when the 
edge was removed. The connecting edge links the two subgraphs together and thus should contribute greatly to 
the algorithmic complexity of the graph. Viewing the connecting edge as a causal entity, meaning that this edge
can only be created once both nodes of which it is connected to are created, then it should be that the 
connecting edge should not only have a positive information contribution but should also contribute highly to 
the algorithmic complexity of the graph. Table~\ref{results_table} shows the effectiveness of edge perturbation
via algorithm $\mathfrak{C}$. For both $\Gamma = \Cup(G)$ and $\Gamma = S_G$ in each graph $G$, 
$\mathfrak{C}(G, \Gamma)$ was mostly able to identify the connecting edge as the edge with the highest average
positive information contribution. 

When using $S_G$ as the parameter for $\mathfrak{C}$, the algorithm was able to identify that the connecting 
edge had the highest average information contribution value for all but one graph compared to $\Cup(G)$ (rows 
colored red in Table~\ref{results_table}). $\Cup(G)$ is already effective in identifying if the connecting edge
is $\maxinfo$ but using $S_G$ was better because it identified the connecting edge as $\maxinfo$ for more 
graphs compared to $\Cup(G)$. The single graph where the algorithm was not successful is the graph where the 
subgraphs were two Erdős–Rényi graphs with 5 vertices each. This should be correct because for most cases of 
random graphs, if two substructures are algorithmically random, then connecting them by a single edge is not 
distinguishable with writing a program that lists all the edges where the connecting edge is also included.

\begin{table}[t!]
\begin{adjustwidth}{-2.25in}{0in}
\centering
\caption{
A list of synthetic graphs where each graph has two subgraphs connected by a single edge.}
\begin{tabular}{ |l|c|c|r|r|c|c| } 
\hline
Connected Graphs ($G$) & $|V_G|$ & $|E_G|$ & $|\Lambda(G)|$ & $|\Lambda(G)|$ / $|S_G|$ & Using $\Lambda(G)$ & 
	Using $S_G$ \\
\thickhline
	Complete4 - Cycle5 & 9 & 12 & 30,240 & 9\% & \cmark & \cmark \\
	Complete5 - Cycle4 & 9 & 15 & 7,560 & 3\% & \cmark & \cmark \\
	Complete4 - Random5 & 9 & 13 & 30,240 & 9\% & \cmark & \cmark \\
	Complete5 - Cycle5 & 10 & 16 & 75,600 & 3\% & \cmark & \cmark \\
	Complete5 - Star5 & 10 & 15 & 25,200 & 1\% & \cmark & \cmark \\
	Complete5 - Complete5 & 10 & 21 & 3,150 & 1\% & \cmark & \cmark \\
	\rowcolor{red!30}
	Star5 - Random5 & 10 & 11 & 604,800 & 17\% & \xmark & \cmark \\
	\rowcolor{red!30}
	Random5 - Random5 & 10 & 13 & 1,814,400 & 50\% & \xmark & \xmark \\
	\rowcolor{red!30}
	Cycle5 - Star5 & 10 & 10 & 302,400 & 9\% & \xmark & \cmark \\
	Cycle4 - Star6 & 10 & 10 & 75,600 & 3\% & \cmark & \cmark \\
	\rowcolor{red!30}
	Cycle5 - Ladder6 & 11 & 13 & 19,958,400 & 50\% & \xmark & \cmark \\
	Cycle5 - Random6 & 11 & 15 & 19,958,400 & 50\% & \cmark & \cmark \\
	Watts-Strogatz6 - Cycle5 & 11 & 18 & 3,326,400 & 9\% & \cmark & \cmark \\
	Complete5 - Random6 & 11 & 20 & 1,663,200 & 5\% & \cmark & \cmark \\
	Cycle5 - Star6 & 11 & 11 & 831,600 & 3\% & \cmark & \cmark \\
	Watts-Strogatz6 - Complete5 & 11 & 23 & 277,200 & 1\% & \cmark & \cmark \\
	Watts-Strogatz6 - Star5 & 11 & 17 & 1,108,800 & 3\% & \cmark & \cmark \\
	Barabási–Albert7 - Complete4 & 11 & 17 & 3,326,400 & 9\% & \cmark & \cmark \\
	Barabási–Albert7 - Cycle4 & 11 & 15 & 9,979,200 & 25\% & \cmark & \cmark \\
	Barabási–Albert6 - Random5 & 11 & 15 & 19,958,400 & 50\% & \cmark & \cmark \\
	Random6 - Random5 & 11 & 16 & 39,916,800 & 100\% & \cmark & \cmark \\
	Complete6 - Complete5 & 11 & 26 & 13,860 & 1\% & \cmark & \cmark \\
	Ladder6 - Random5 & 11 & 14 & 19,958,400 & 50\% & \cmark & \cmark \\
	Ladder6 - Complete5 & 11 & 18 & 831,600 & 3\% & \cmark & \cmark \\
	\rowcolor{red!30}
	Ladder6 - Star5 & 11 & 12 & 3,326,400 & 9\% & \xmark & \cmark \\
	Watts-Strogatz6 - Random6 & 12 & 22 & 79,833,600 & 17\% & \cmark & \cmark \\
	Watts-Strogatz7 - Complete5 & 12 & 25 & 19,958,400 & 5\% & \cmark & \cmark \\
	Watts-Strogatz7 - Star5 & 12 & 19 & 79,833,600 & 17\% & \cmark & \cmark \\
	Barabási–Albert7 - Star5 & 12 & 15 & 39,916,800 & 9\% & \cmark & \cmark \\
	Complete6 - Complete6 & 12 & 31 & 16,632 & 1\% & \cmark & \cmark \\
\hline
\end{tabular}
\begin{flushleft}  
	The first column shows the names of the two subgraphs that were connected and how many vertices they have,
	e.g. Cycle5 - Star6 means a cycle graph with 5 vertices and a star graph with 6 vertices. The second 
	and third column shows the number of vertices $V_G$, number of edges $E_G$ of each graph respectively. 
	The forth column displays how many automorphic subsets $|\Lambda(G)|$ each graph has, and the
	fifth column is the portion of the amount of automorphic subsets within its symmetric group. The higher the
	percentage, the more random the graph is, since having many automorphic subsets means the graph is not 
compressible to a limited number of permutations. The last two columns shows a check mark if 
	$\mathfrak{C}$ was able to determine if the connecting edge has the highest average information contribution
	in graph $G$ using: (a) one permutation for each automorphic subset, $\mathfrak{C}(G, \Cup(G))$ (sixth
	column) and (b) all permutations from the entire symmetric group $\mathfrak{C}(G, S_G)$ (last column). The
	red rows show the graphs whose edge with the highest average information contribution is not the connecting
	edge when using one permutation from each automorphic subset (marked with an x mark in the sixth column).
\end{flushleft}
\label{results_table}
\end{adjustwidth}
\end{table}

\subsection*{Edge Grouping}

To further determine the effectiveness of edge perturbation, we also analyzed how similar the information 
contribution of an edge to other edges. Since we have computed the average information contribution for every 
edge of a graph $G$ using $\mathfrak{C}$, we can then compare if edges per sub-graph ($G_1$ \& $G_2$) have 
grouped correctly using the sorted set of tuples $I$ (recall that $I$ is arranged from highest to lowest in 
average information contribution value). When $\mathfrak{C}$ outputs $I$, it shows the likelihood of each edge
to be grouped with other edges that were produced by the same underlying sub-program. We categorize each graph 
into three different grouping schemes with respect to the linear arrangement of the average information 
contribution values of $E_G$:

\begin{enumerate}
	\item {Complete - all the edges in $I$ have grouped $G_1$ and $G_2$'s edges respectively.}
	\item {Partial - 60\% or more of either $G_1$ or $G_2$'s edges grouped together.}
	\item {Scattered - no edges of either $G_1$ or $G_2$ were grouped together with at least 60\% from the 
		same sub-graph.}
\end{enumerate} Examples of these grouping schemes can be seen in Figure \nameref{S1_Fig}.

Averaging of information contribution values has been done because different vertex permutations result in 
different $K(G)$ values. Since $\Lambda(G)$ groups isomorphisms of different vertex permutations that are
automorphic per grouping, $\Cup(G)$ is the compressed version of $\Lambda(G)$ and ultimately of $S_G$. Pairing 
this approach of getting unique permutations and averaging of each edge's information contribution in a graph 
allows us to have a fair assessment of $BDM$ by probabilistic and combinatorial means. We computed the average 
information contribution values with respect to the parameter $\Gamma$ in $\mathfrak{C}(G, \Gamma)$ to
determine if $\Cup(G)$ is already effective without using $S_G$ since $\Cup(G)$ is algorithmically quicker to 
execute. Tables \ref{automorphic_groupings_table} and \ref{symmetric_groupings_table} show a comparison of 
$\Cup(G)$ and $S_G$. We can see that the symmetric group is more effective compared to using automorphic subset 
sampling. Although using $\Cup(G)$ has partially grouped edges of some graphs, $S_G$ correctly grouped edges 
for more graphs. 

\begin{table}[t!]
\begin{adjustwidth}{-2.25in}{0in}
\small
\centering
\caption{Edge grouping schemes using an element in each automorphic subset of each graph.}
\begin{tabular}{|ccc|}
	\hline
	\multicolumn{3}{|c|}{$\Gamma = \Cup(G)$} \\
	\hline
	Complete & Partial & Scattered \\
	\thickhline
	Complete5 - Cycle4 & Barabási–Albert7 - Cycle4 & Barabási–Albert6 - Random5 \\
	Watts-Strogatz7 - Complete5 & Barabási–Albert7 - Star5 & Barabási–Albert7 - Complete4 \\
	 & Complete4 - Cycle5 & Complete5 - Complete5 \\
	 & Complete4 - Random5 & Complete6 - Complete5 \\
	 & Complete5 - Cycle5 & Complete6 - Complete6 \\
	 & Complete5 - Random6 & Cycle5 - Ladder6 \\
	 & Complete5 - Star5 & Cycle5 - Star5 \\
	 & Cycle4 - Star6 & Ladder6 - Random5 \\
	 & Cycle5 - Random6 & Ladder6 - Star5 \\
	 & Cycle5 - Star6 & Random5 - Random5 \\
	 & Ladder6 - Complete5 & Star5 - Random5 \\
	 & Random6 - Random5 & Watts-Strogatz6 - Complete5 \\
	 & Watts-Strogatz6 - Cycle5 & \\
	 & Watts-Strogatz6 - Random6 & \\
	 & Watts-Strogatz6 - Star5 & \\
	 & Watts-Strogatz7 - Star5 & \\
	 \hline
\end{tabular}
\label{automorphic_groupings_table}
\end{adjustwidth}
\end{table}

\begin{table}[t!]
\begin{adjustwidth}{-2.25in}{0in}
\small
\centering
\caption{Edge grouping schemes using the symmetric group of each graph.}
\begin{tabular}{|ccc|}
	\hline
	\multicolumn{3}{|c|}{$\Gamma = S_G$} \\
	\hline
	Complete & Partial & Scattered \\
	\thickhline
		Barabási–Albert7 - Complete4 & Barabási–Albert7 - Star5 & Barabási–Albert6 - Random5 \\
		Barabási–Albert7 - Cycle4 & Complete4 - Cycle5 & Complete4 - Random5 \\
		Complete5 - Cycle4 & Complete5 - Complete5$^*$ & Cycle5 - Ladder6 \\
		Complete5 - Cycle5 & Complete6 - Complete5$^*$ & Random5 - Random5 \\
		Complete5 - Random6 & Complete6- Complete6$^*$ & Random6 - Random5 \\
		Complete5 - Star5 & Cycle4 - Star6 & \\
		Cycle5 - Star6 & Cycle5 - Random6$^*$ & \\
		Ladder6 - Complete5 & Cycle5 - Star5$^*$ & \\
		Ladder6 - Star5 & Ladder6 - Random5 & \\
		Watts-Strogatz7 - Complete5 & Star5 - Random5 & \\
		& Watts-Strogatz6 - Complete5$^*$ & \\
		& Watts-Strogatz6 - Cycle5 & \\
		& Watts-Strogatz6 - Random6 & \\
		& Watts-Strogatz6 - Star5 & \\
		& Watts-Strogatz7 - Star5 & \\
	\hline
\end{tabular}
\begin{flushleft}  
	$^*$Partial grouping has been achieved, but with a more refined grouping where at least two edges of a 
	sub-graph cluster together to form a group.
\end{flushleft}
\label{symmetric_groupings_table}
\end{adjustwidth}
\end{table}

\begin{figure}[H]
  \includegraphics[width=0.85\textwidth]{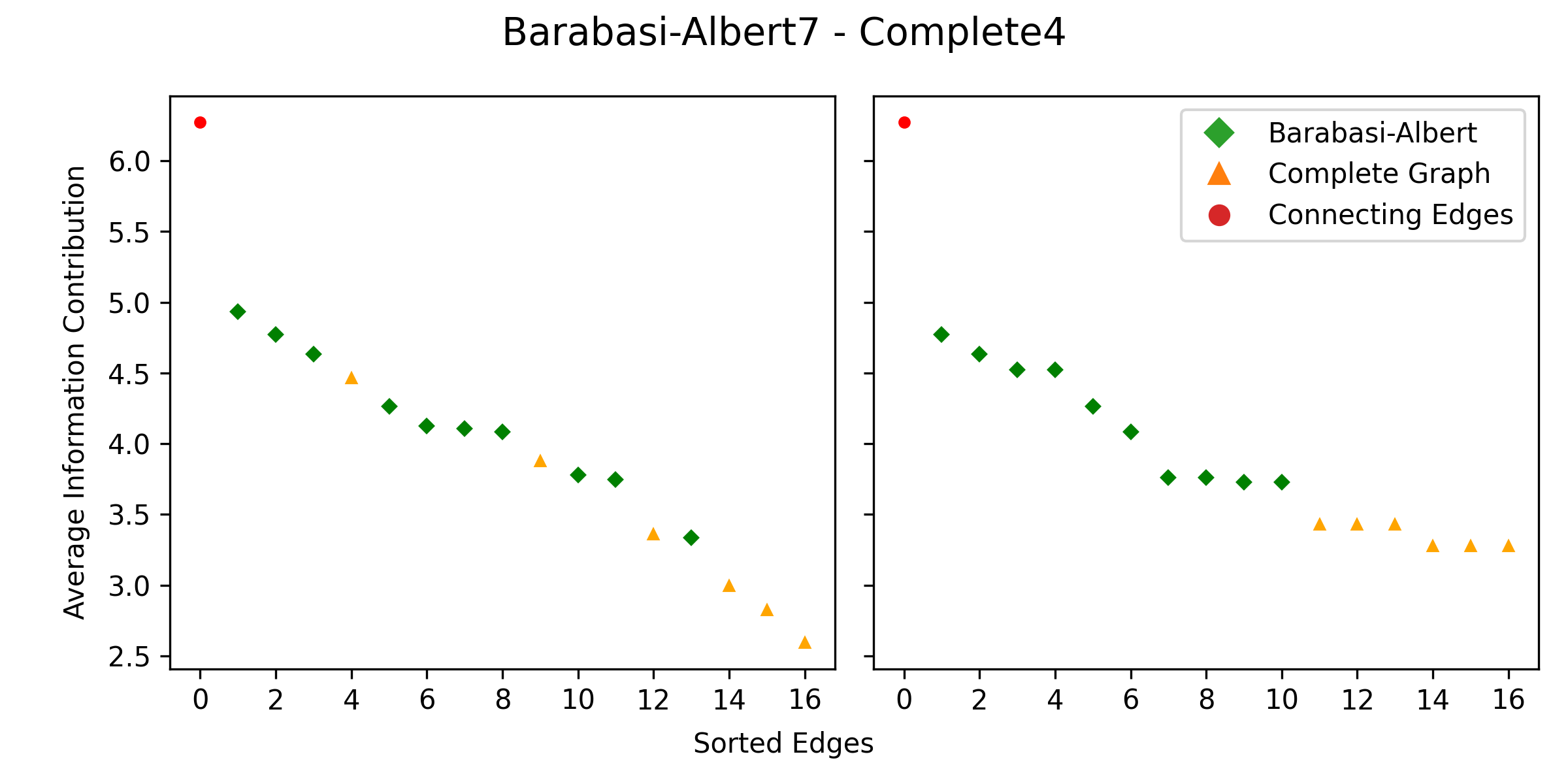}
  	\\[0.5em]
  \includegraphics[width=0.85\textwidth]{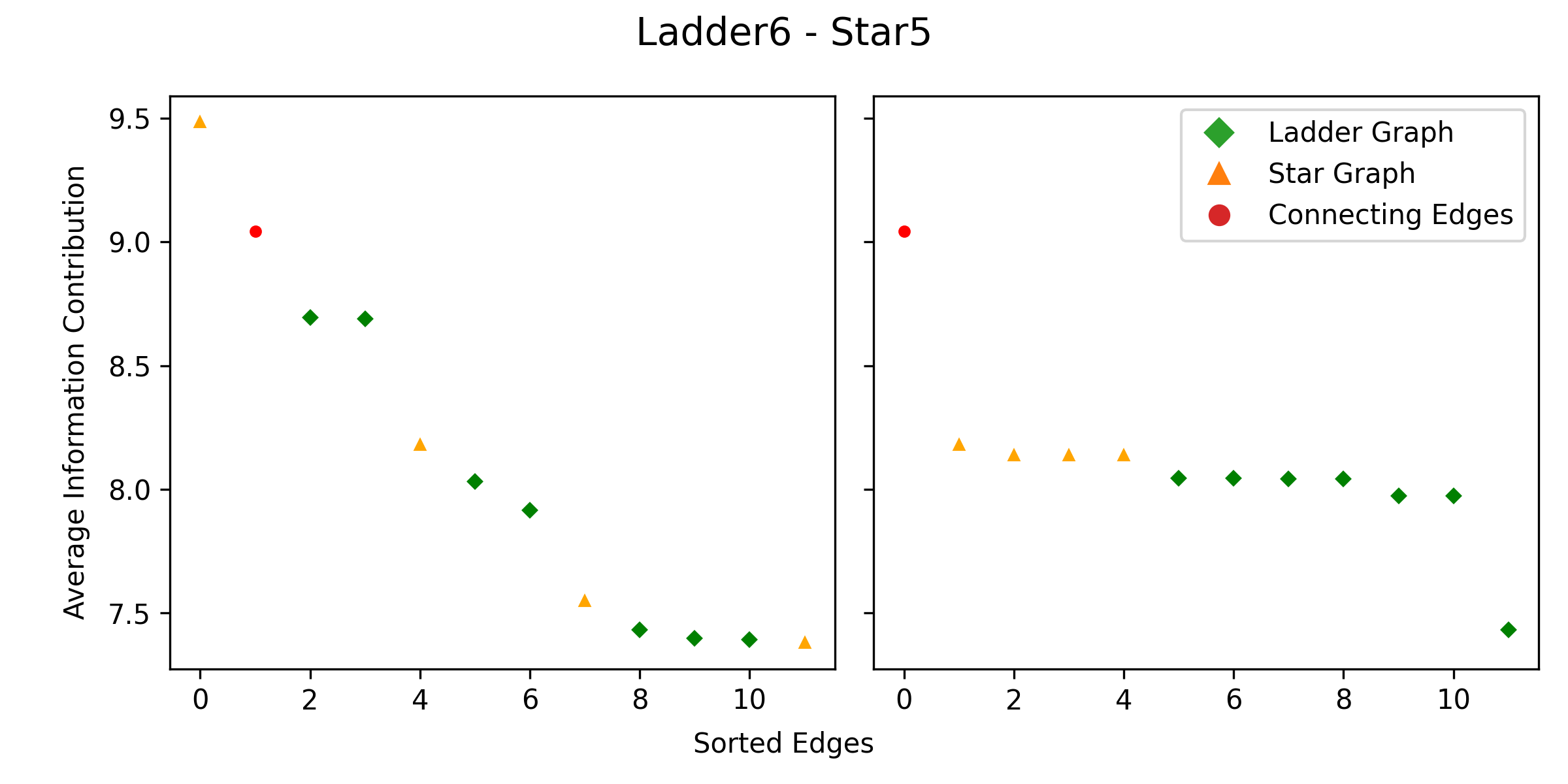}
  \caption{Graphs whose initial edges grouping was \textbf{scattered} when using
  	$\Gamma = \Cup(G)$(left) but was classified as complete when $\Gamma = S_G$(right).}
  \label{fig:scattered_to_complete}
\end{figure}

\begin{figure}[H]
  \includegraphics[width=0.85\textwidth]{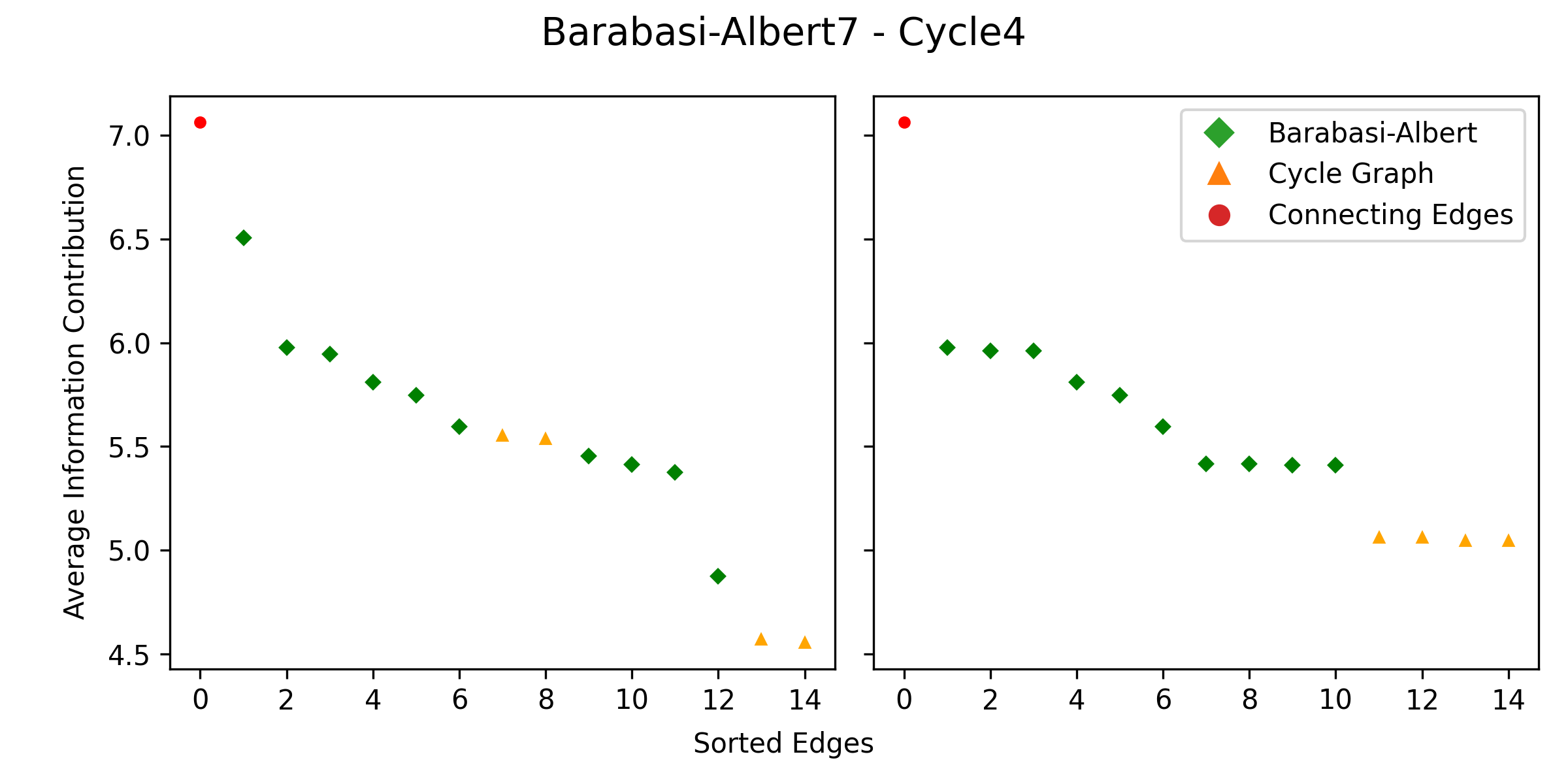}\\[0.5em]
  \includegraphics[width=0.85\textwidth]{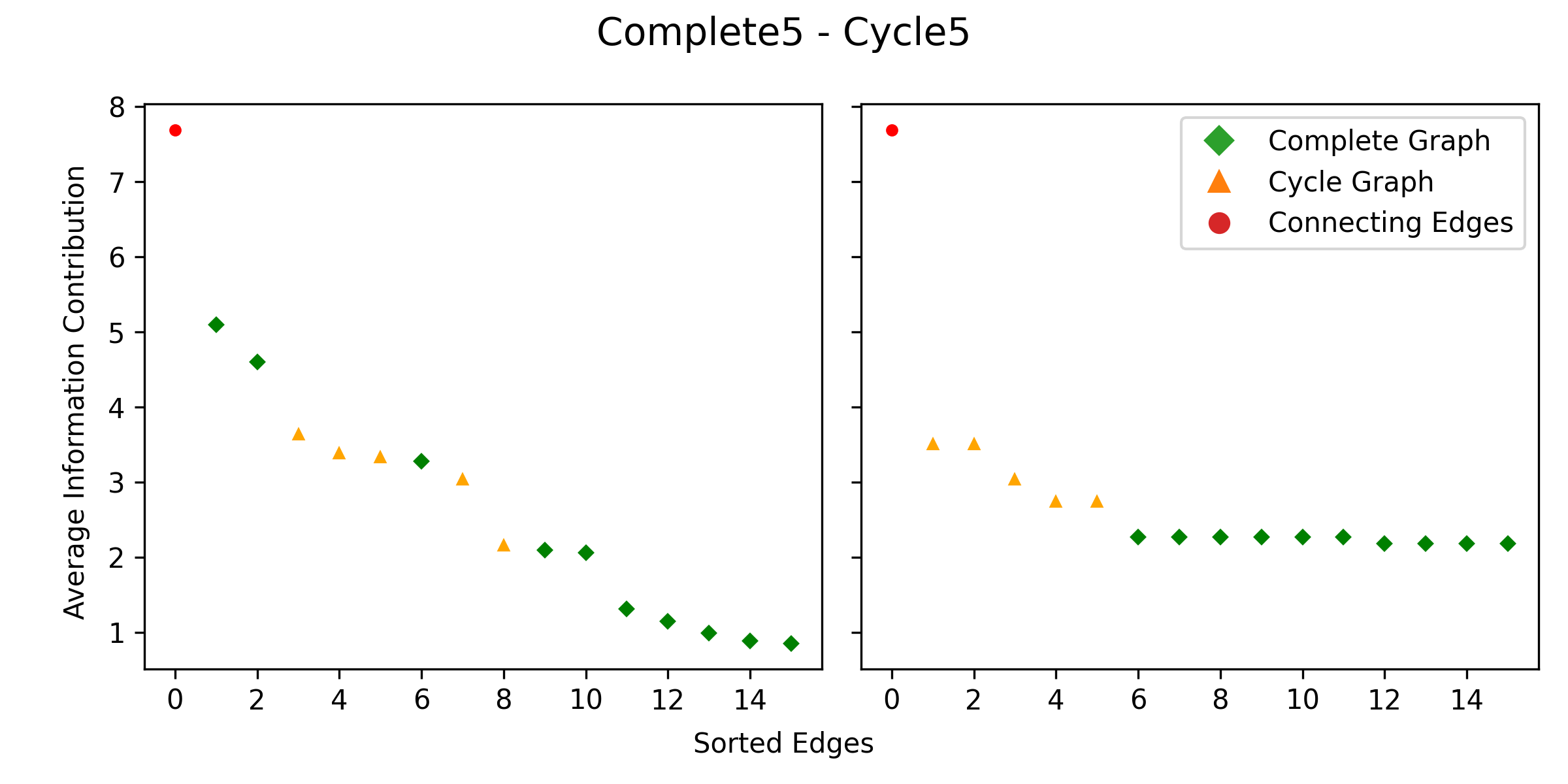}\\[0.5em]
  \includegraphics[width=0.85\textwidth]{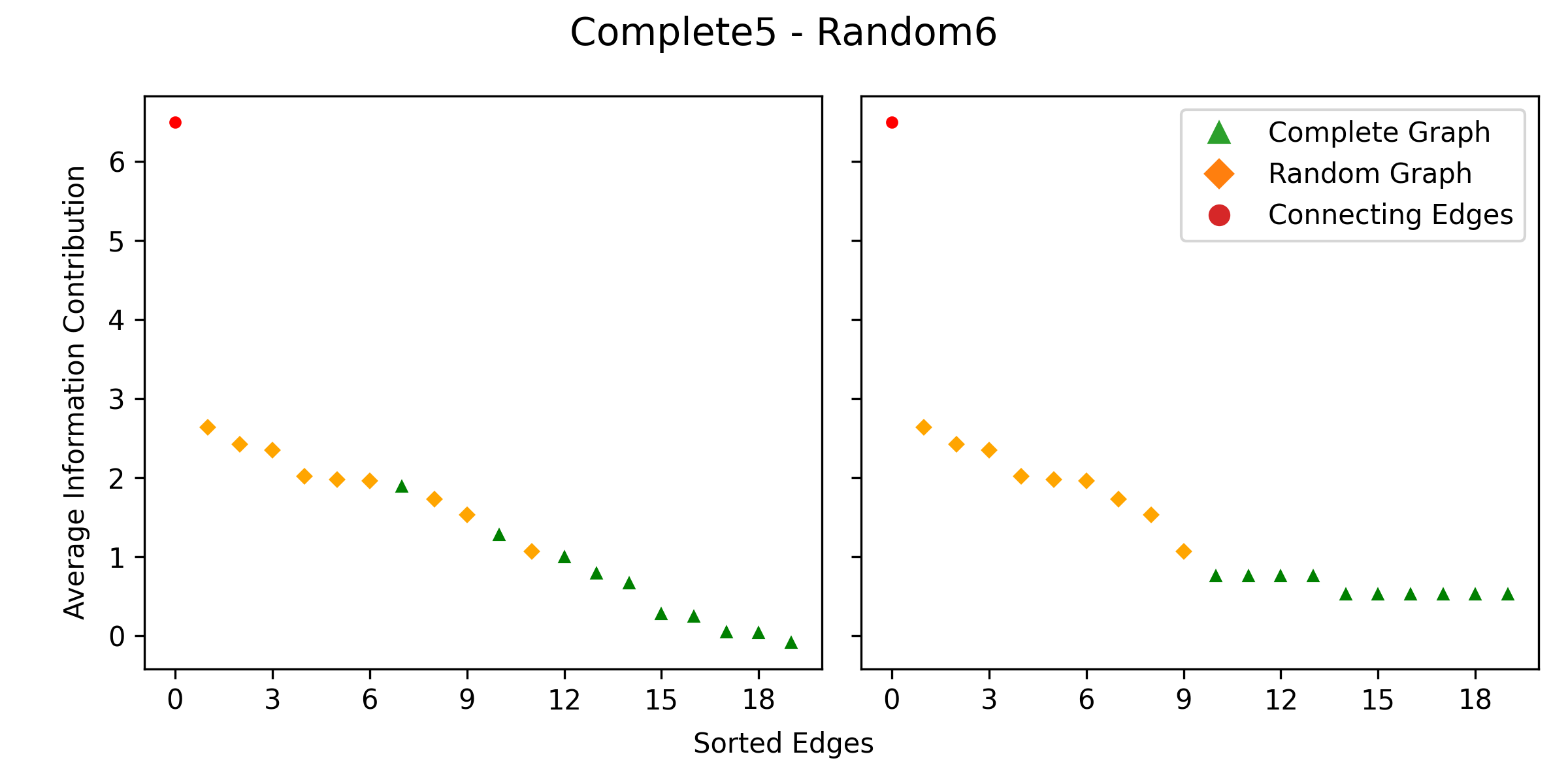}
  	\caption{Graphs whose initial edges grouping was \textbf{partial} when using
  	$\Gamma = \Cup(G)$(left) but was classified as complete when $\Gamma = S_G$(right).}
  \label{fig:partial_to_complete1}
\end{figure}

\begin{figure}[H]
  \includegraphics[width=0.85\textwidth]{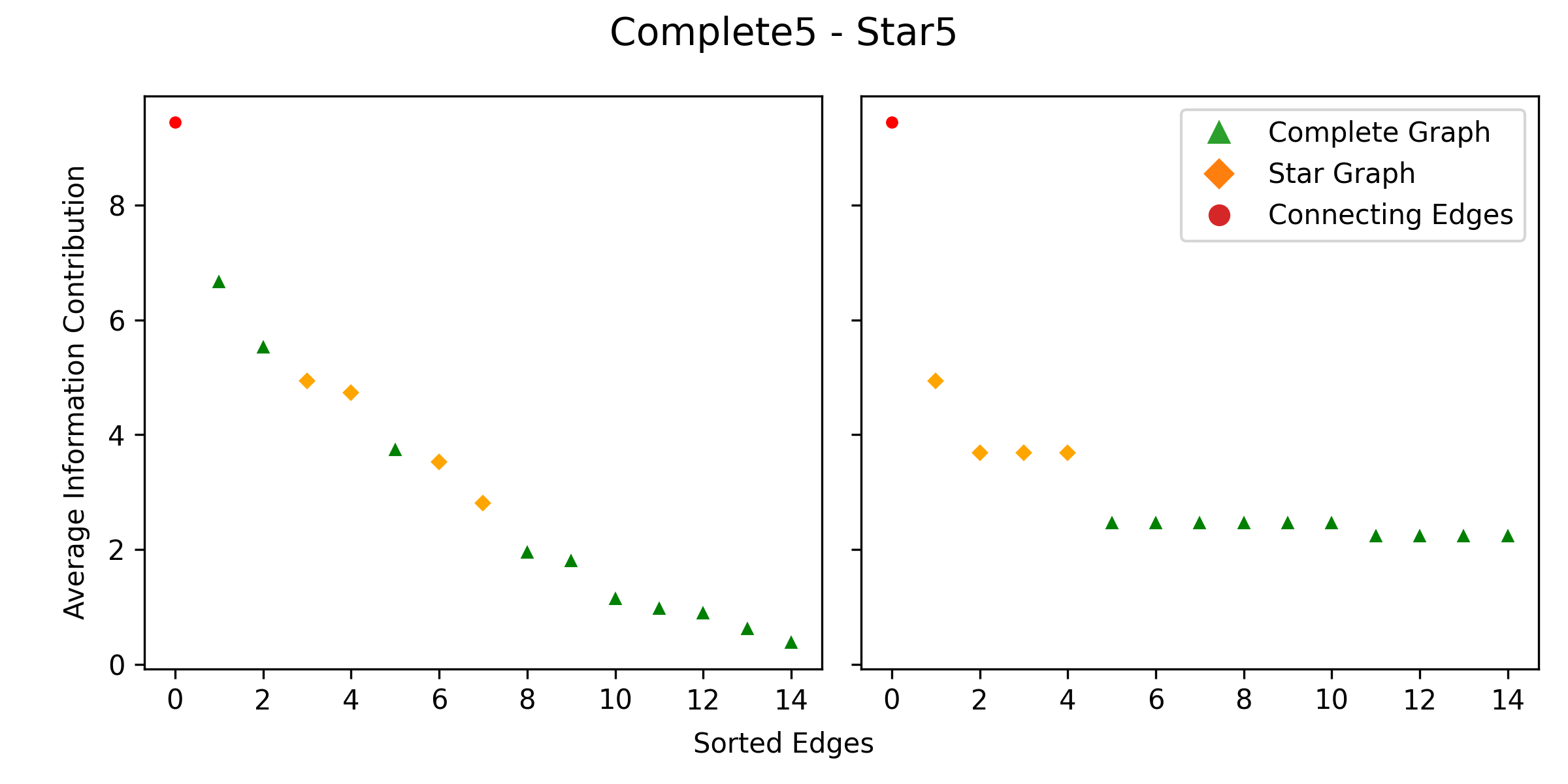}\\[0.5em]
  \includegraphics[width=0.85\textwidth]{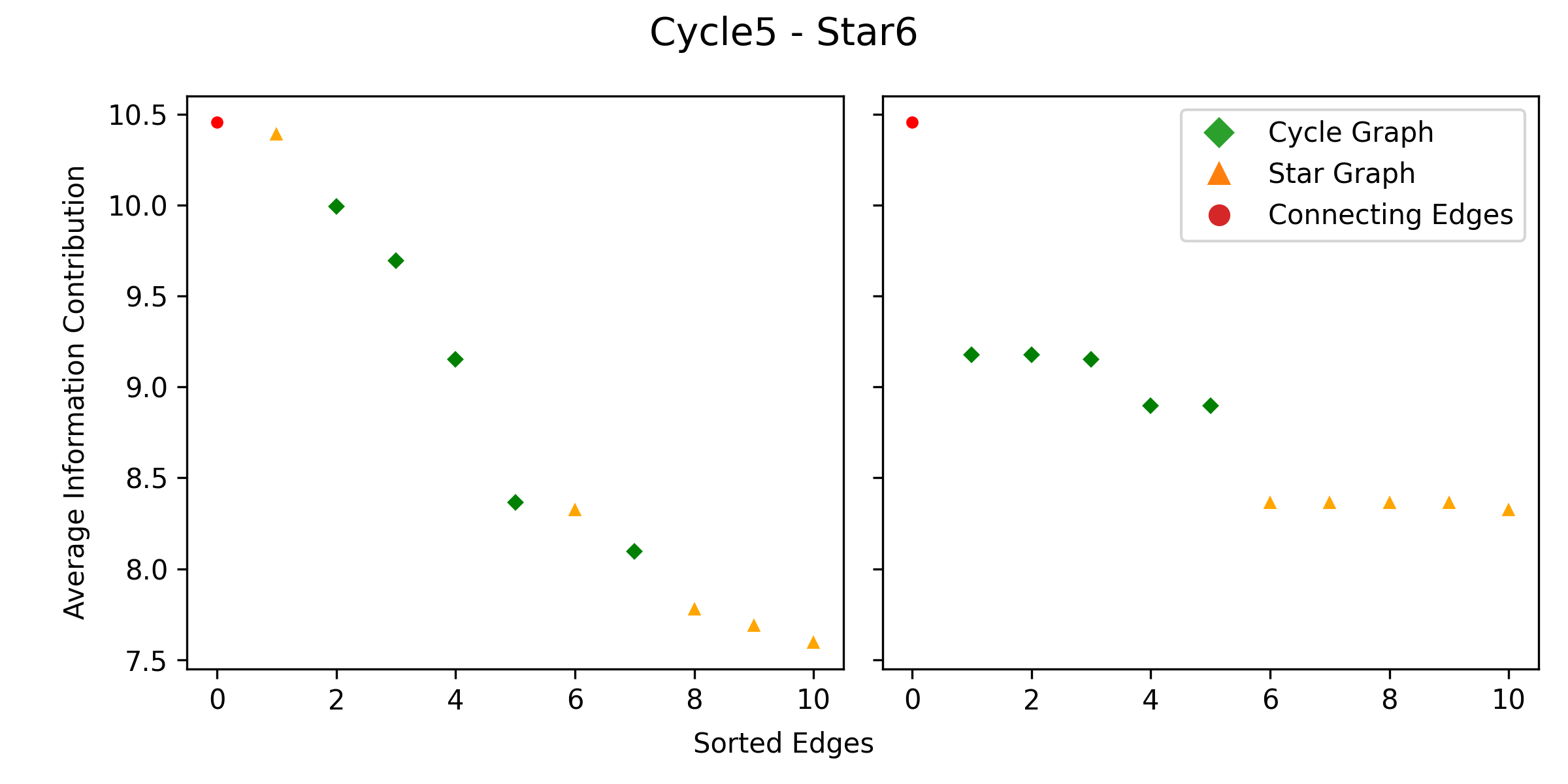}\\[0.5em]
  \includegraphics[width=0.85\textwidth]{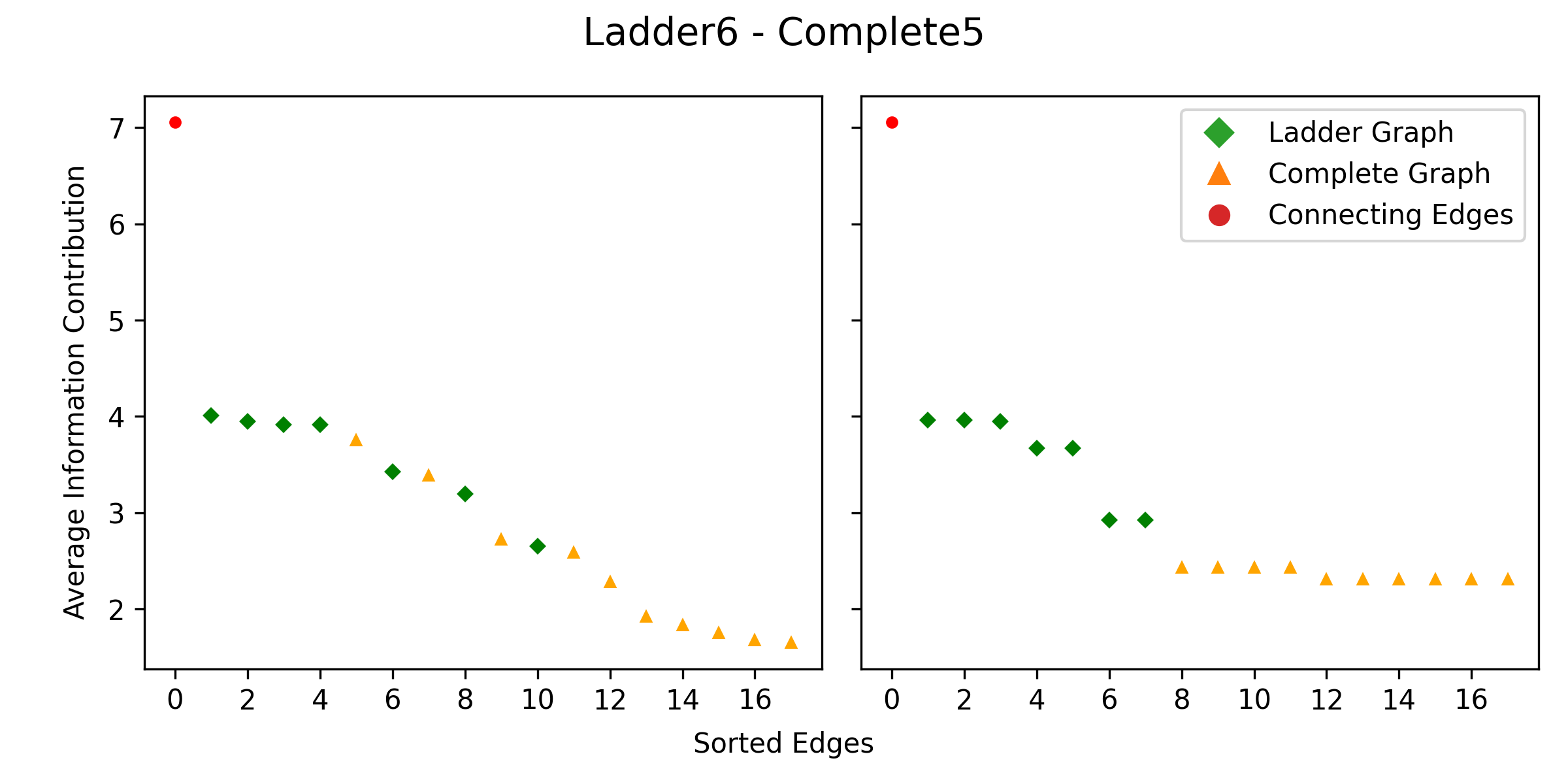}
  \caption{Graphs whose initial edges grouping was \textbf{partial} when using
  	$\Gamma = \Cup(G)$(left) but was classified as complete when $\Gamma = S_G$(right).}
  \label{fig:partial_to_complete2}
\end{figure}

\begin{figure}[H]
  \includegraphics[width=0.85\textwidth]{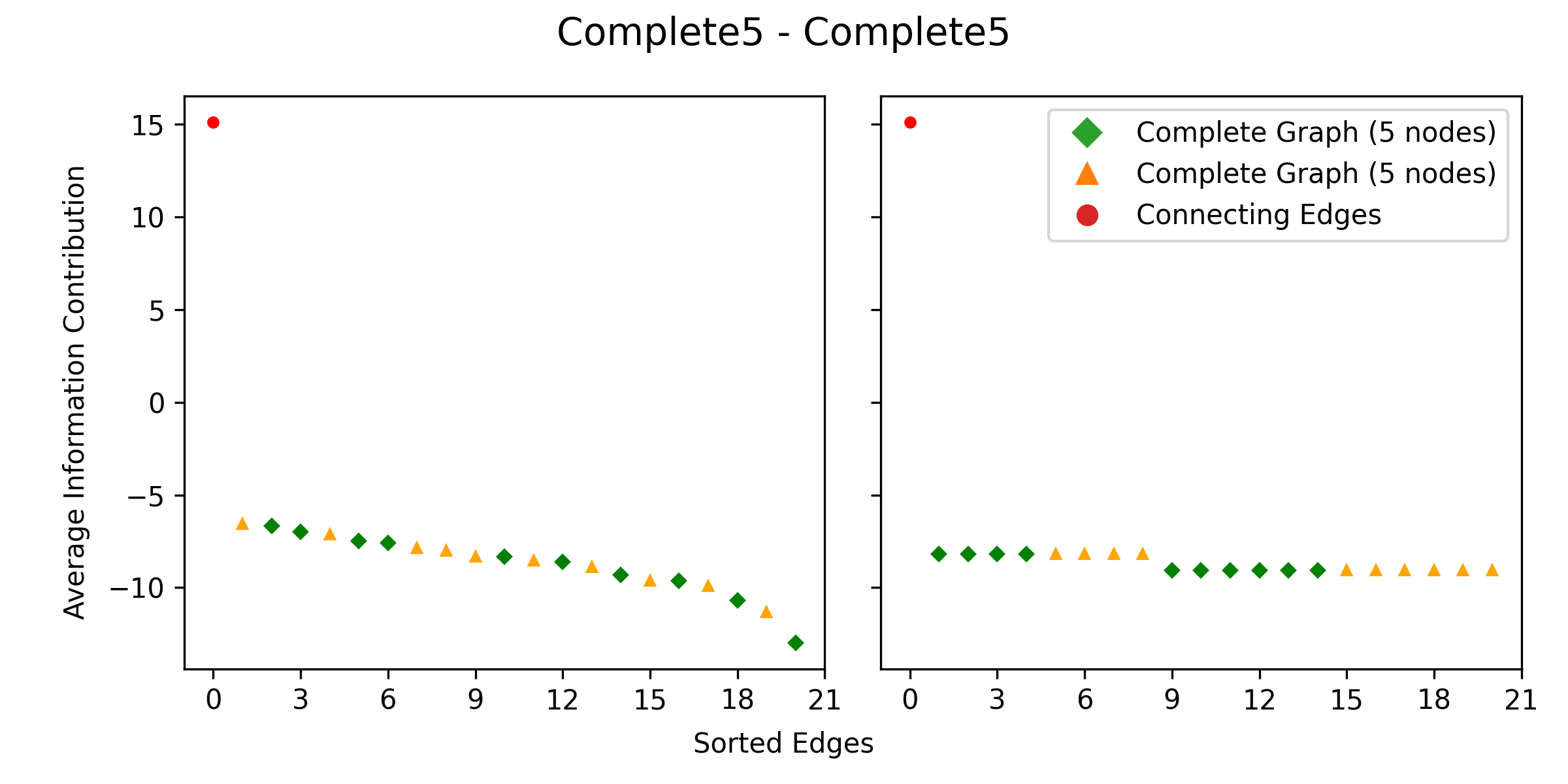}\\[0.5em]
  \includegraphics[width=0.85\textwidth]{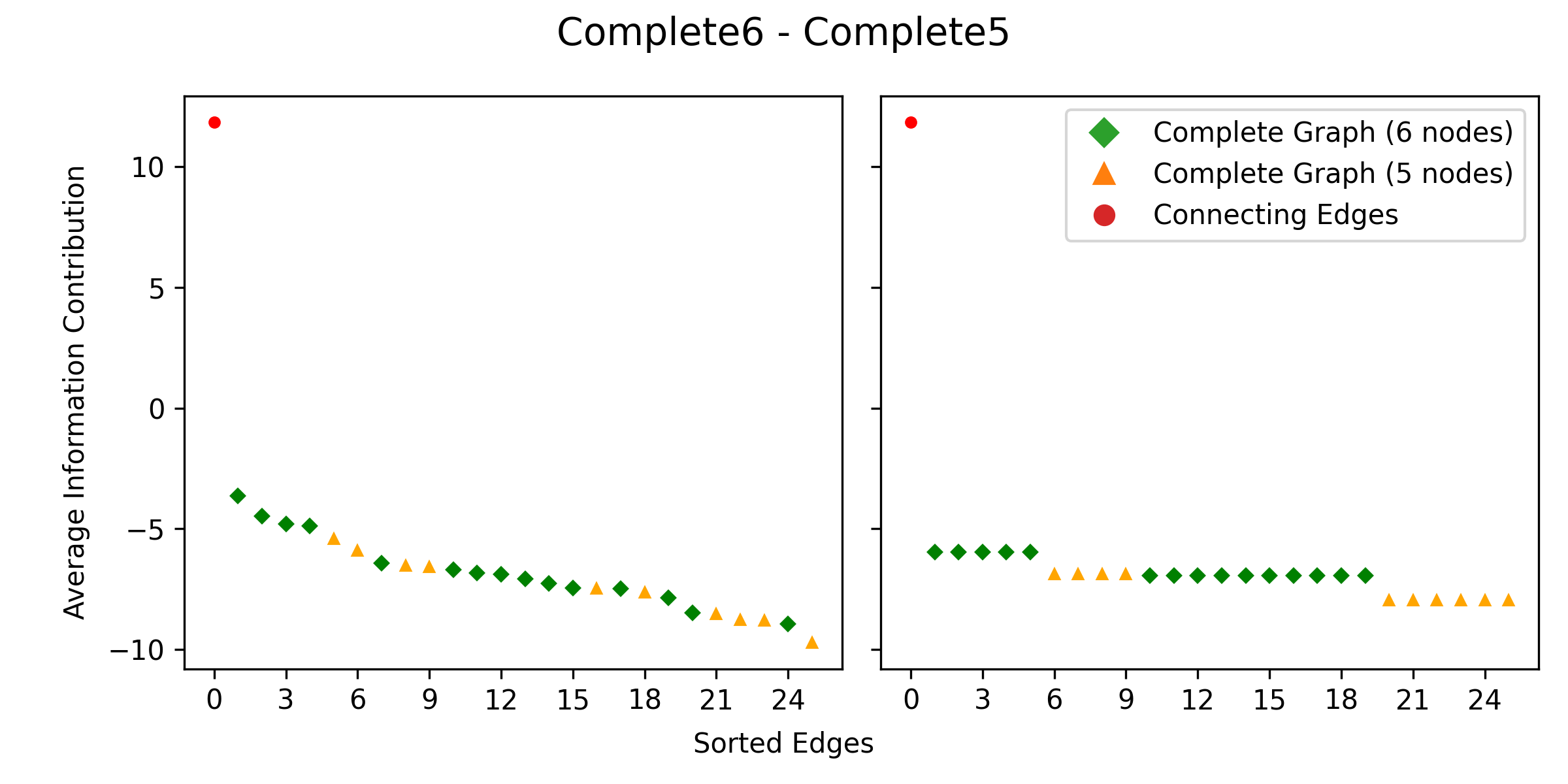}\\[0.5em]
  \includegraphics[width=0.85\textwidth]{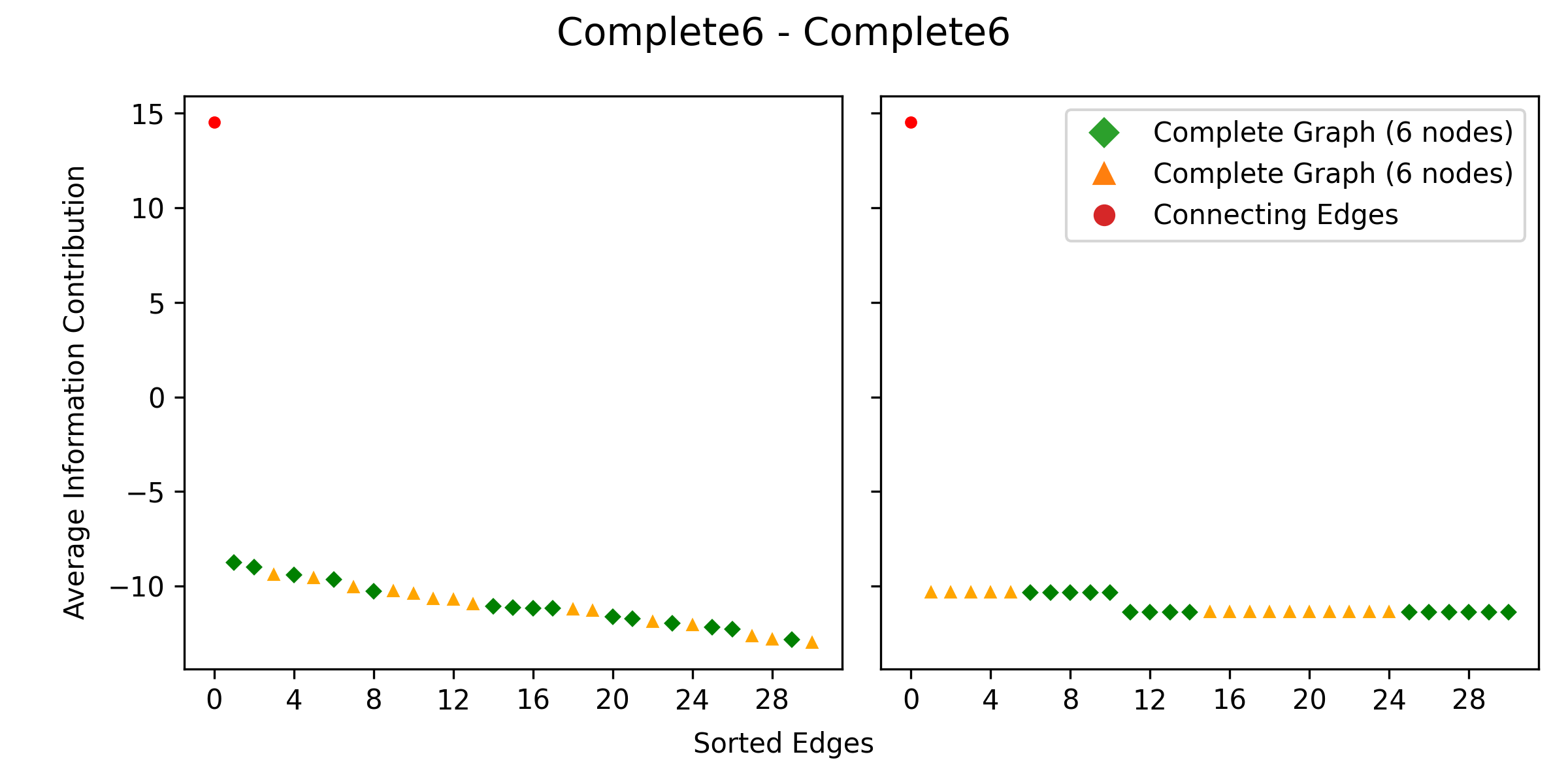}
    \caption{Graphs that are categorized as scattered or partial edge grouping when
		$\Gamma = \Cup(G)$(left) but was refined to have a better grouping when $\Gamma = S_G$(right) was 
		used.}
  \label{fig:refined_partial1}
\end{figure}

\begin{figure}[H]
  \includegraphics[width=0.85\textwidth]{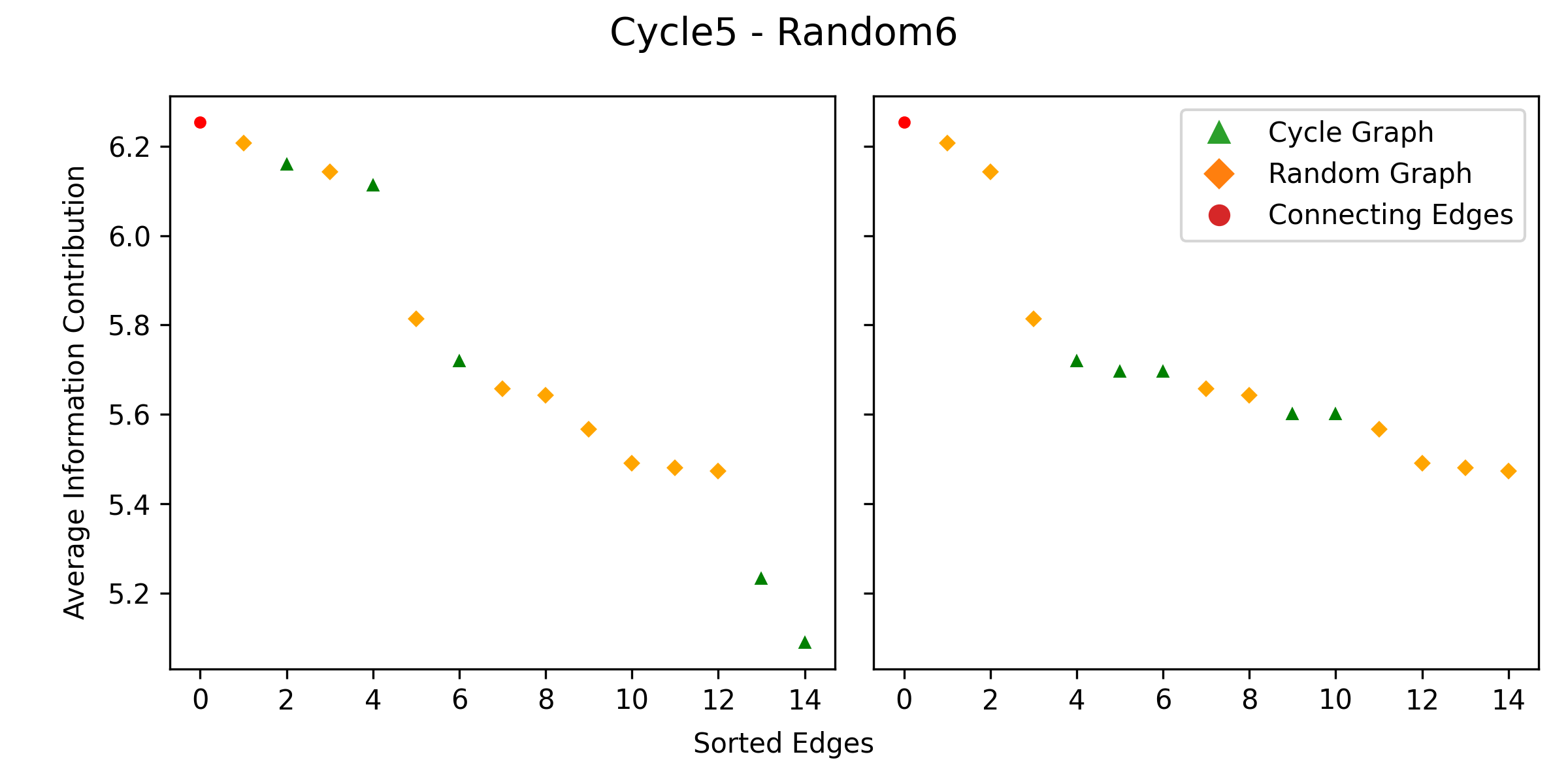}\\[0.5em]
  \includegraphics[width=0.85\textwidth]{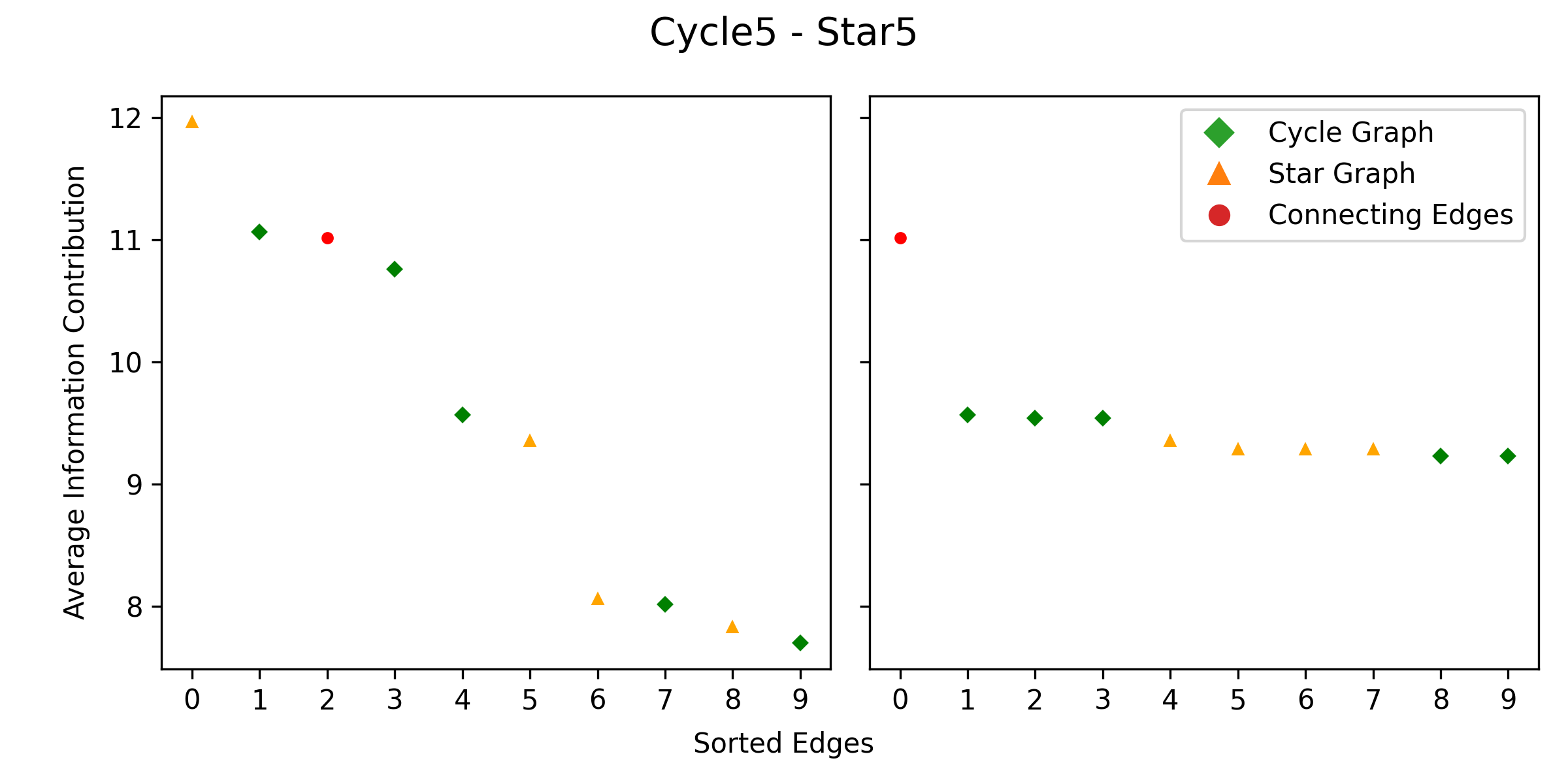}\\[0.5em]
  \includegraphics[width=0.85\textwidth]{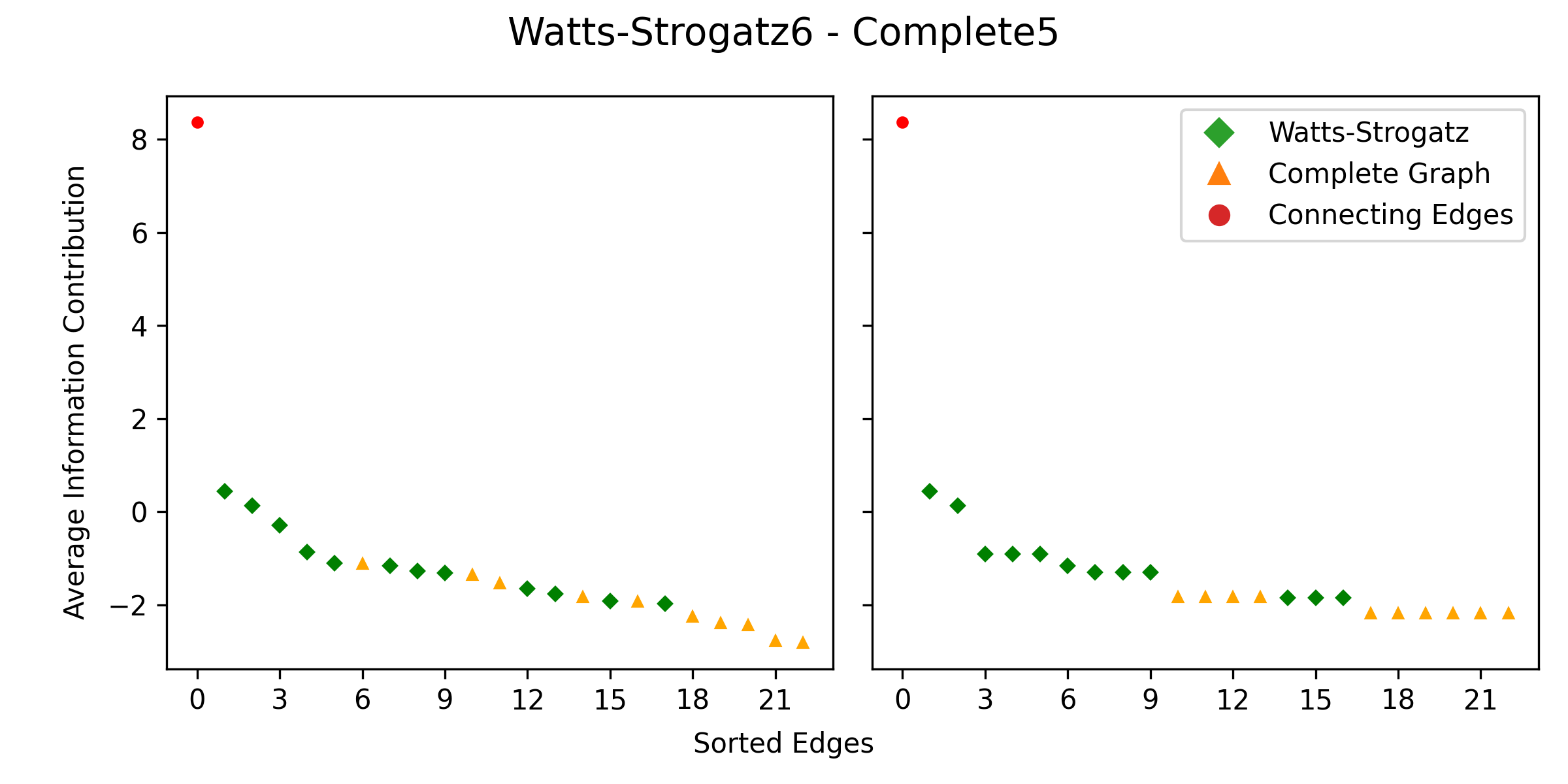}
  \caption{Graphs that are categorized as scattered or partial edge grouping when
		$\Gamma = \Cup(G)$(left) but was refined to have a better grouping when $\Gamma = S_G$(right) was 
		used.}
  \label{fig:refined_partial2}
\end{figure}

A side-by-side comparison in Figures~\ref{fig:scattered_to_complete} to~\ref{fig:refined_partial2} 
highlights further the effectiveness of using $S_G$. To highlight, we can also observe that some partially 
grouped edges in some graphs have shown improvements in terms of their linear arrangement when $\Gamma = S_G$ 
over $\Gamma = \Cup(G)$. These graphs are marked with $^*$ in Table~\ref{symmetric_groupings_table} and their 
improved information contribution grouping can be seen in Figures~\ref{fig:refined_partial1} and 
~\ref{fig:refined_partial2}. 

\subsection*{Information Contribution Distance}

Finally, edge perturbation not only tries to determine how much information does an edge contribute to the 
overall complexity of a graph but also shows the causal relation of each edge to the entire graph. 
Since the connecting edge is causally dependent on the existence of $G_1$ and $G_2$ of a graph $G$, it is 
theoretically expected to have a distance greater than $\log_2(2)$ from the other edges for most graphs 
(assuming it is not a totally random graph). In Equation~\ref{eq:algocomplexitytwographs}, the connecting edge
$e'$ is included in $O(1)$. If $v_{G_1}$ and $v_{G_2}$ are the vertices (each belong to each sub-graph) that 
are connected by $e'$, then $G_1^*$ and $G_2^*$ should be executed / instantiated first so that $e'$ can 
connect the two subgraphs $G_1$ and $G_2$. The presence of $e'$ increases the algorithmic complexity of the 
graph in terms of not only program size, but also causal relations from previously executed sub-programs when 
instantiating the entirety of $G$.

Using a combination of the information contribution of an edge and the growth of program lengths, we can 
determine which edges are instantiated by which sub-programs in $G^*$ (the shortest program that produces $G$). 
In general, the growth of program lengths is $\log_2(2)$~\cite{bib8}, this implies that when performing edge 
perturbation on edge $e_i$ and $e_j$ (for example), where both edges are included in the same sub-graph, their 
information contribution values should have a distance of not more than $\log_2(2)$ from one another. 
This is because that there are exponentially shorter programs to describe a program than there are to describe
longer ones, thus it is exponentially unlikely that two edges with a difference of $\log_2(2)$ will be 
generated by the same program.

\begin{table}[t!]
\begin{adjustwidth}{-2.25in}{0in}
\small
\centering
\caption{The information contribution distance of the connecting edge for each
	of the 30 graphs tested.}
\begin{tabular}{|c|l|l|}
	\hline
	\multirow{2}{*}{Connected Graphs ($G$)} & \multicolumn{2}{c|}{$e_n - e_{n-1}$} \\
	\cline{2-3} & \multicolumn{1}{c|}{$\Gamma = \Cup(G)$} & \multicolumn{1}{c|}{$\Gamma = S_G$} \\
	\thickhline
	\rowcolor{red!30}
	Barabasi-Albert6 - Random5 & 0.13378793460520733 & 0.13378793486272755 \\
	Barabasi-Albert7 - Complete4 & 1.3376391875417948 & 1.502790069936819 \\
	\rowcolor{red!30}
	Barabasi-Albert7 - Cycle4 & 0.5571016664781094 & 1.0879645196663352 \\
	\rowcolor{red!30}
	Barabasi-Albert7 - Star5 & 0.5638933790503007 & 0.5638933794518453 \\
	Complete4 - Cycle5 & 2.8631256102071503 & 3.172163330027856 \\
	Complete4 - Random5 & 1.3977474661420275 & 1.3977474661462583 \\
	Complete5 - Complete5 & 21.6912793586853 & 23.317882074217472 \\
	Complete5 - Cycle4 & 5.317601412837774 & 6.002081335753412 \\
	Complete5 - Cycle5 & 2.595326258259141 & 4.183456312163415 \\
	Complete5 - Random6 & 3.857727731304253 & 3.857727730900918 \\
	Complete5 - Star5 & 2.792869861977472 & 4.512949749130205 \\
	Complete6 - Complete5 & 15.485835510254873 & 17.817978351324065 \\
	Complete6 - Complete6 & 23.27621702970346 & 24.880830306329372 \\
	\rowcolor{red!30}
	Cycle4 - Star6 & 0.4198132053221766 & 3.6647074566079105 \\
	\rowcolor{red!30}
	Cycle5 - Ladder6 & 0.06675395441072052$^\dagger$ & 0.19422595474726378 \\
	\rowcolor{red!30}
	Cycle5 - Random6 & 0.04686230316845297 & 0.04686230326476615 \\
	\rowcolor{red!30}
	Cycle5 - Star5 & 0.25489223766848035$^\dagger$ & 1.4481659979161563 \\
	\rowcolor{red!30}
	Cycle5 - Star6 & 0.06969126690328409 & 1.2786870792130305 \\
	Ladder6 - Complete5 & 3.0520261513245543 & 3.099060700177053 \\
	\rowcolor{red!30}
	Ladder6 - Random5 & 0.39179993216664855 & 0.4665223320742333 \\
	\rowcolor{red!30}
	Ladder6 - Star5 & 0.34964872859838714$^\dagger$ & 0.8657191703469174 \\
	\rowcolor{red!30}
	Random5 - Random5 & 0.027091365649044796$^\dagger$ & 0.14635893537649114$^\dagger$ \\
	\rowcolor{red!30}
	Random6 - Random5 & 0.22022158873730113 & 0.22022158867519082 \\
	\rowcolor{red!30}
	Star5 - Random5 & 0.5541784484738965$^\dagger$ & 0.5541784484676509 \\
	Watts-Strogatz6 - Complete5 & 7.936125911627891 & 7.936125911614662 \\
	Watts-Strogatz6 - Cycle5 & 1.5279233787798026 & 1.6305962768312012 \\
	Watts-Strogatz6 - Random6 & 2.6902761749130186 & 2.690276175861212 \\
	Watts-Strogatz6 - Star5 & 1.4985545797543267 & 1.5750285631331034 \\
	Watts-Strogatz7 - Complete5 & 4.317408673610543 & 4.317408676679262 \\
	\rowcolor{red!30}
	Watts-Strogatz7 - Star5 & 0.9944753991092607 & 0.9944753983040977 \\
	\hline
\end{tabular}
\begin{flushleft}
	$^\dagger$The connecting edge does not have $\max_{\text{info}}$ using this permutation set. Let $e_n$ be 
	the connecting edge which is placed in the $n$th location in $I_k = \{ e_{n+1}, e_n, e_{n-1},
	\ldots, e_1 \}$, where $I_k$ is the set of keys for the set of tuples $I$ and $I[e_n]$ is the average 
	information contribution value of the connecting edge. If the connecting edge is $\max_{\text{info}}$, then
	$n = |I|$ (the first value in the linear arrangement).
\end{flushleft}
\label{difference_less_log2_table}
\end{adjustwidth}
\end{table}

As seen in Table~\ref{difference_less_log2_table}, not only did $\mathfrak{C}$ identify that the connecting 
edge was $\maxinfo$ for most graphs using both $\Gamma = \Cup(G)$ and $\Gamma = S_G$, but it also managed to 
identify the distance of the connecting edge to the next edge in the linear arrangement of average information 
contribution was more than the cutoff for half of the graphs. Out of the 30 graphs, 16 of them satisfied 
$e_n - e_{n-1} > \log_2(2)$. Interestingly, it shows that performing averaging of information contribution with
either $\Cup(G)$ or $S_G$ was effective when graphs contain a sub-graph that has a high edge count. Among these 
16 graphs, either a Watts-Strogatz or a Complete graph was present as a sub-graph for each graph. This 
indicates that $BDM$ is effective in detecting tightly connected substructures within a graph. Further proof 
is that when a graph only contains these two subgraphs, the connecting edge's average information contribution
is high relative to the other graphs examined.

Conversely, $BDM$ did not satisfy $e_n - e_{n-1} > \log_2(2)$ for 14 out of the 30 graphs (colored red in 
Table~\ref{difference_less_log2_table}) when $\Gamma = \Cup(G)$. 6 of the 9 graphs have a random sub-graph that
is included per graph.  When both subgraphs are random graphs, then adding a connecting edge will not be 
distinguishable from the other edges in the graph because the entire graph would be considered one combined 
random graph. There was not much change in $e_n - e_{n-1}$ when using $S_G$ over $\Cup(G)$ except for graphs 
that had a cycle sub-graph. This indicates that averaging of information contribution using $S_G$ provides a
more fine-grained inspection of substructures. Appending to the findings in the Edge Grouping section of the 
results, most of the graphs that were classified as complete or partial edge grouping consist of mostly graphs
with regular and small-world subgraphs, namely Cycle, Complete and Watts-Strogatz graphs.

\section*{Discussion}

There are 2 limitations that prohibit us from further experimentation of larger graphs. First is that 
$\mathfrak{C}$ grows by $n!$, where $n$ is the number of nodes of the graph. To fully test $BDM$ on the basis
of combinatorial and probabilistic means, much bigger graphs should be tested using $\mathfrak{C}$. Although
$\mathfrak{C}$ shows promise of identifying substructures in a graph and was able to: 

\begin{itemize}
	\item Determine that the connecting edge was $\maxinfo$ except for 1 graph using $S_G$
	\item Completely (10 graphs) or partially (15 graphs) group edges based on their average information 
		contribution in a linear arrangement using $S_G$
	\item Differentiate the connecting edge by having a distance of more than $\log_2(2)$ from the other edges
		in 20 graphs when using $S_G$
\end{itemize}

it must be tested in real world networks whose number of nodes is greater than 12 to have a more accurate
testing of $BDM$. The second limitation is in the sub-matrix size that $CTM$ (within $BDM$) can slice an 
adjacency matrix of a graph. Currently, $BDM$ can handle $3 \times 3$ or $4 \times 4$ sub matrices and uses the 
periodic partitioning when $n$ is not divisible by $l$. Ideally, since $BDM$ is an approximated upper-bound of
algorithmic complexity, we would like for $CTM$ to handle $l > 4$ for larger graphs since there is an error 
overhead whenever $n$ is not divisible by $l$ and thus the remaining rows and columns in the adjacency matrix 
are appended with the previous rows and columns using a partitioning technique. But this is a monumental task 
since computing for larger matrices in $CTM$ is equivalent to running the Busy Beaver function.

On a tangential topic of recursively applying edge perturbation. We can determine if an edge can be cut from 
a graph (to reveal substructures) when an edge fits the two criteria that we have discussed in the results, 
which are:

\begin{itemize}
	\item[--] the edge needs to have a high information contribution compared to the other edges
	\item[--] the edge should have a distance of at least $\log_2(2)$ 
\end{itemize}

If these criteria have been satisfied by an edge, then we can remove it from the graph. The graphs that have 
satisfied these criteria are the 16 graphs in Table~\ref{difference_less_log2_table} (rows not colored red). By
removing edges in the graph we can recursively apply edge perturbation to the new graph since the adjacency 
matrix of the new graph has been altered and thus different information contribution values will be assigned to
the remaining edges. Using the new information contribution values, we can determine which edges can be removed
along with edge grouping (using the linear arrangement of these values) to identify substructures. The more
connecting edges are removed, the more edge grouping becomes accurate. For example, if a graph $G$ has three
subgraphs namely, $G_1, G_2, G_3$. If there are edges that connect the subgraphs with one another, then they 
are removed using edge perturbation recursively until they have all been removed, the algorithmic complexity of 
$K(G) \leq |G_1^*| + |G_2^*| + |G_3^*|$. This allows $\mathfrak{C}$ to approximate the edges that belong to the
subgraphs (which are sub-programs of $G$) only and no added sub-programs are needed since there are no 
remaining connecting edges. If $m$ is the number of recursions we apply edge perturbation and $n$ is the number
of nodes in the graph, then the runtime of $\mathfrak{C}$ with respect to $m$ is $mn!$. Although this is still
$\mathcal{O}(n!)$, computing $\mathfrak{C}$ becomes more expensive when recursion is applied.

\section*{Conclusion}

A combination of permuting a graph then averaging the information contribution per edge perturbed was done to 
30 synthetic graphs to determine the efficacy of $BDM$ as a method in revealing substructures within complex 
networks. The sets of permutations that the algorithm $\mathfrak{C}$ (which uses $BDM$) used was the 
automorphic subsets $\Cup(G)$ of a graph and its symmetric group $S_G$. We used these two permutation sets to 
compare a compressed representation of a graph's structure (automorphic subsets) from its entire permutation
set (symmetric group). Out of the 30 graphs, 29 of them have been successfully identified as having the 
connecting edge as the edge with the highest average information contribution. Within these 29 graphs, 16 were 
also identified of having the connecting edge as the edge with a distance of more than $\log_2(2)$ from other 
edges within their respective graphs. To highlight, edge perturbation was proven to be effective for graphs 
that have tightly connected subgraphs, namely Watts-Strogatz graphs and Complete graphs. The connecting edge 
was clearly differentiated in terms of its average information contribution from the edges of the subgraphs. 
Averaging of information contribution using $\Cup(G)$ and $S_G$ was also effective in completely or partially 
grouping edges together. Comparing the two permutation sets, $S_G$ performed better since more graphs' edge 
grouping has classified as a complete edge grouping.

There are several improvements / suggestions we offer to extend experimentation with this methodology.
One avenue is to add more edges when connecting one or more subgraphs. We have only connected the 
two subgraphs by one edge to have a simple testing scenario. By adding more edges or increasing the number 
of subgraphs, we can have varying results since the graphs are more diverse and may also mimic real-life 
networks. A second avenue is instead of using automorphic subsets or the symmetric group of the graph, 
a random sample of permutations with a fixed size of $10!$ can be used so that regardless of the size of the 
graph, we can have constant time in testing a high enough number of graphs whose number of nodes is greater 
than 12. And a last avenue is how we assign information contribution values to an edge. In this paper, we have 
used averaging as a means of assignment, other possible assignment options can be the standard deviation of the
information contribution of a perturbed edge from the average algorithmic complexity of $S_G$ or a smaller
permutation set. Another option is getting the minimum algorithmic complexity in $S_G$ then relating that to
a perturbed edge's information contribution when its graph $G$ is permuted by $\sigma \in S_G$.

\newpage 

\section*{Supporting information}

\paragraph*{S1 Fig.}
\label{S1_Fig}
{\bf Group scheming for graphs }(using $\Gamma = \Cup(G)$). First: Complete and Cycle graph 
	(Complete), Second: Watt-Strogatz and Cycle (Partial), Last: Barabási–Albert and Erdős–Rényi (Scattered).
\begin{figure}[H]
	\includegraphics[width=0.58\textwidth]{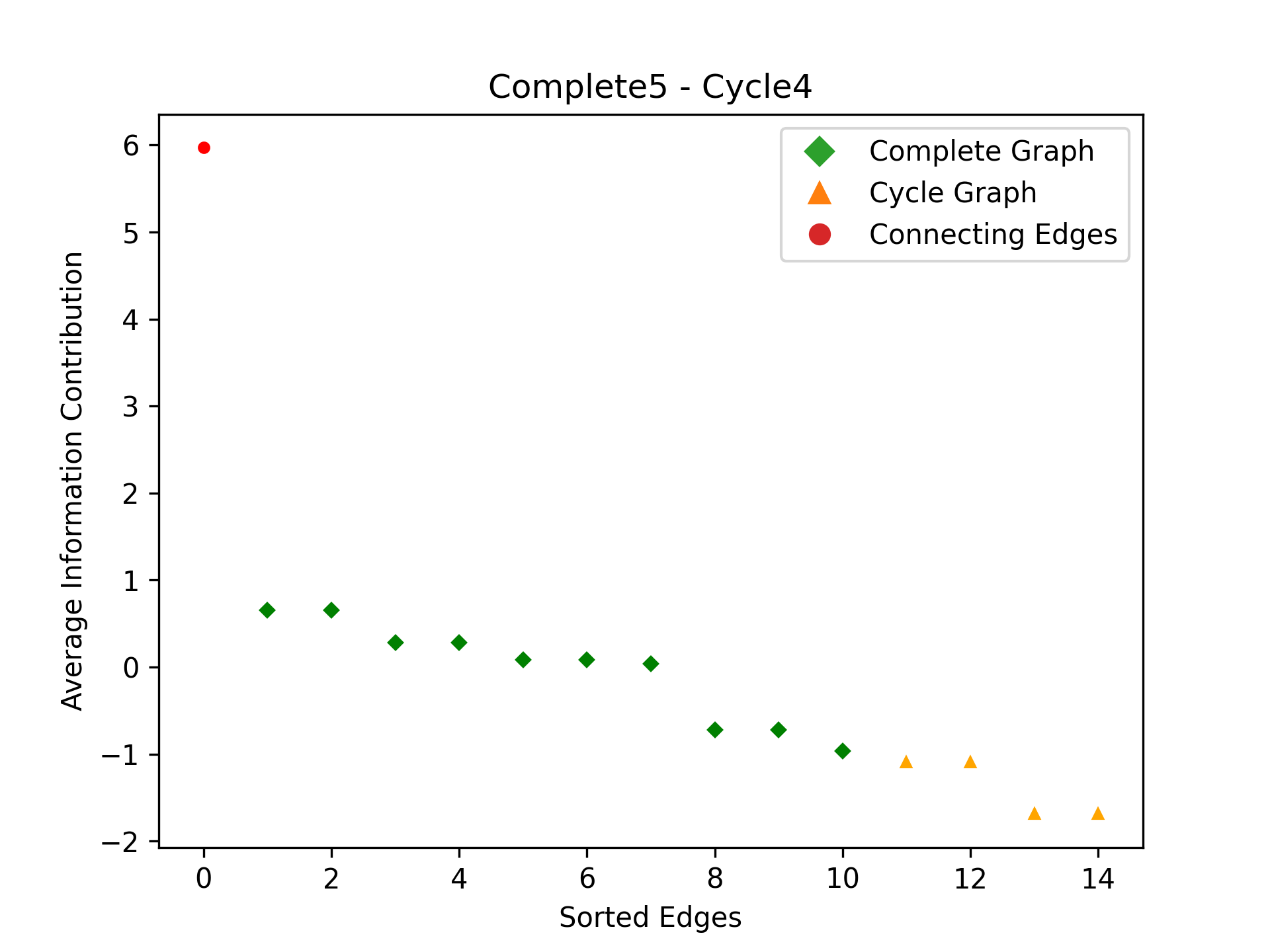}\\[0.5em]
    \includegraphics[width=0.58\textwidth]{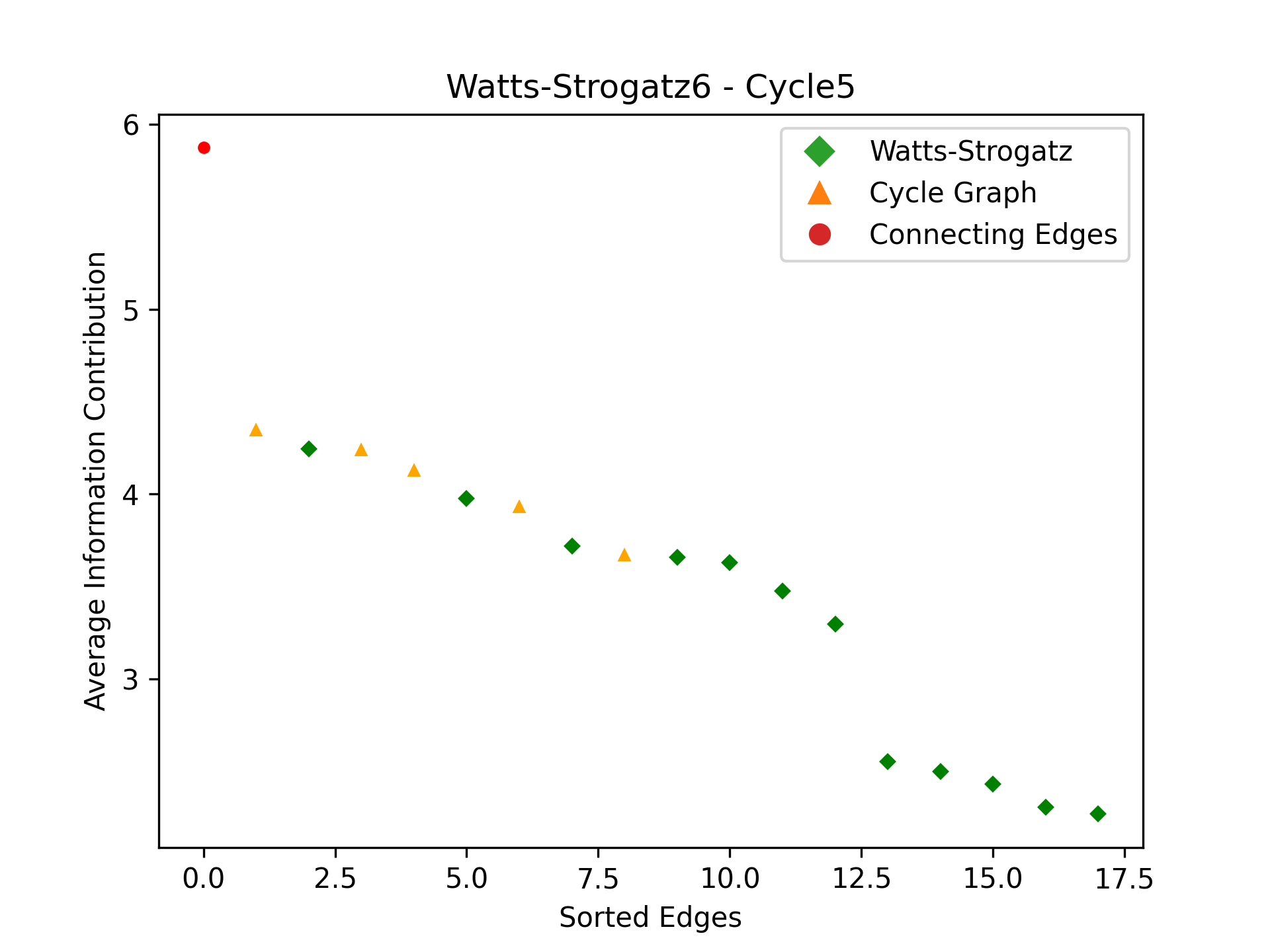}\\[0.5em]
    \includegraphics[width=0.58\textwidth]{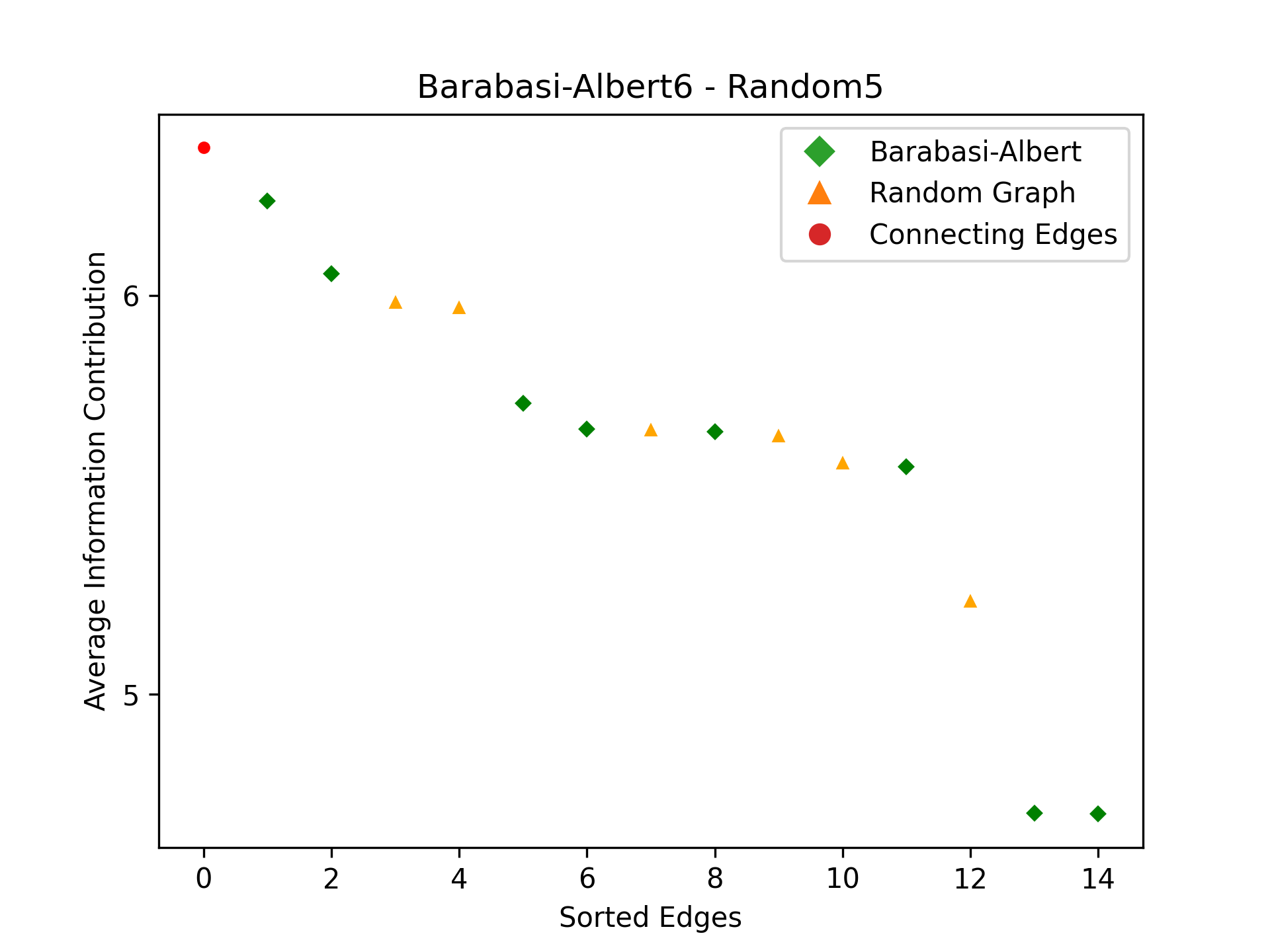}
\end{figure}



\section*{Acknowledgments}
I would like to acknowledge my thesis adviser Briane Samson for guiding and motivating me when writing this 
paper. I would also like to thank Hector Zenil for answering my inquiries regarding their initial 
implementation of causal deconvolution along with Allan Zea, who implemented the R package of causal 
deconvolution. Lastly, I would like to show my gratitude to the maintainer of the Python package that 
implements $BDM$, Szymon Talaga, for answering questions regarding implementing edge perturbation to the 
package.

\nolinenumbers

%
%
%


\begin{thebibliography}{20}

\bibitem{bib1}
Mitchell, M.
\newblock {{C}omplexity: A Guided Tour}.
\newblock OUP USA, 2011.

\bibitem{bib2}
Rossi, R., Ahmed, N.
\newblock {{T}he Network Data Repository with Interactive Graph Analytics and Visualization}.
\newblock AAAI, \url{https://networkrepository.com}, 2015. 

\bibitem{bib3}
Estrada, E.
\newblock {{W}hat is a complex system, after all?}.
\newblock Foundations of Science, 2023, 29(4), 1143–1170 

\bibitem{bib4}
Rabin, M. O.
\newblock {{T}uring Centennial Conference: Turing, Church, Gödel, Computability, Complexity and Randomization}.
\newblock Google TechTalks. (2012, April 25), YouTube. \url{https://www.youtube.com/watch?v=ofyXXOpRB0U}

\bibitem{bib5}
Zenil, H., Hernández-Orozco, S., Kiani, N. A., Soler-Toscano, F., Rueda-Toicen, A., \& Tegnér, J.
\newblock {{A} Decomposition Method for Global Evaluation of Shannon Entropy and Local Estimations of 
Algorithmic Complexity}.
\newblock Entropy, 2018, 20(8), 605.

\bibitem{bib6}
Soler-Toscano, F., Zenil, H., Delahaye, J., \& Gauvrit, N.
\newblock {{C}alculating Kolmogorov Complexity from the Output Frequency Distributions of Small Turing 
Machines}.
\newblock PLoS ONE, 2014, 9(5), e96223.

\bibitem{bib7}
Zenil, H., Soler-Toscano, F., Delahaye, J., \& Gauvrit, N.
\newblock {{T}wo-dimensional Kolmogorov complexity and an empirical validation of the Coding theorem method by
compressibility}.
\newblock Journal of the ACM, 1975, 22(3), 329–340.

\bibitem{bib8}
Zenil, H., Kiani, N. A., Zea, A. A., \& Tegnér, J.
\newblock {{C}ausal deconvolution by algorithmic generative models}.
\newblock Nature Machine Intelligence, 2018, 1(1), 58–66.

\bibitem{bib9}
Zenil, H., Kiani, N. A., Marabita, F., Deng, Y., Elias, S., Schmidt, A., Ball, G., \& Tegnér, J.
\newblock {{A}n Algorithmic Information Calculus for Causal Discovery and Reprogramming Systems}.
\newblock iScience, 2019, 19, 1160–1172.

\bibitem{bib10}
McKay, B. \& Piperno, D.
\newblock {{P}ractical Graph Isomorphism, II}
\newblock J. Symbolic Computation, 2013, 60, 94-112.

\bibitem{bib11}
Chaitin, G.
\newblock {{A} theory of program size formally identical to information theory}.
\newblock Journal of the ACM, 1975, 22(3), 329–340.

\bibitem{bib12}
Kolmogorov, A.
\newblock {{T}hree approaches to the quantitative definition of information}.
\newblock International Journal of Computer Mathematics, 1968, 2(1–4), 157–168.

\bibitem{bib13}
Solomonoff, R.
\newblock {{A} formal theory of inductive inference. Part I}.
\newblock Information and Control, 1964, 7(1), 1–22.

\bibitem{bib14}
Levin, L.
\newblock {{L}aws of Information Conservation (Nongrowth) and Aspects of the Foundation of Probability Theory}.
\newblock Problems Inform. Transmission, 1974, 10:3, 206–210

\bibitem{bib15}
Cover, T. \& Thomas, J.
\newblock {{E}lements of Information Theory}.
\newblock John Wiley \& Sons, 2005.

\bibitem{bib16}
Radó, T.
\newblock {{O}n non-computable functions}.
\newblock Bell System Technical Journal, 1962, 41, pp. 877-884.

\bibitem{bib17}
Hartke, S., \& Radcliffe, A.
\newblock {McKay\'s Canonical Graph Labeling Algorithm}
\newblock Contemporary Mathematics - American Mathematical Society, 2009, 99–111.

\bibitem{bib18}
Huaylla, C. A., Kuperman, M. N., \& Garibaldi, L. A.
\newblock {Comparison of two statistical measures of complexity applied to ecological bipartite networks}
\newblock 2024, Physica a Statistical Mechanics and Its Applications, 642, 129764.

\bibitem{bib19}
Dobosan, P.
\newblock {Python Implementation of Nauty}.
\newblock 2022, [Software]. Github. \url{https://github.com/pdobsan/pynauty}

\bibitem{bib20}
Talaga, S., \& Tsampourakis, K.
\newblock PyBDM: Python interface to the Block Decomposition Method (0.1.0) 
\newblock 2024, [Software]. \url{https://zenodo.org/doi/10.5281/zenodo.10652064}

\end{thebibliography}
\end{document}